\makeatletter
\@ifundefined{@parse@version@dash}{%
	\def\@parse@version#1{\@parse@version@0#1}
	\def\@parse@version@#1/#2/#3#4#5\@nil{%
		\@parse@version@dash#1-#2-#3#4\@nil}
	\def\@parse@version@dash#1-#2-#3#4#5\@nil{%
		\if\relax#2\relax\else#1\fi#2#3#4 }
}{}
\makeatother

\documentclass[aps,pre,twocolumn,groupedaddress,longbibliography]{revtex4-2}
\usepackage{amsfonts,amssymb,amsmath} 
\usepackage{color} 
\usepackage{graphicx} 
\usepackage{epstopdf,mathrsfs} 
\usepackage[colorlinks,linktocpage,linkcolor=blue]{hyperref}

\begin{document}

\title{Random diffusivity processes in an external force field}
\author{Xudong Wang$^1$}
\author{Yao Chen$^2$}
\email{ychen@njau.edu.cn}
\affiliation{$^1$School of Mathematics and Statistics, Nanjing University of Science and Technology, Nanjing, 210094, P.R. China \\
$^2$College of Sciences, Nanjing Agricultural University, Nanjing, 210094, P.R. China}

\begin{abstract}
Brownian yet non-Gaussian processes have recently been observed in numerous biological systems and the corresponding theories have been built based on random diffusivity models. Considering the particularity of random diffusivity, this paper studies the effect of an external force acting on two kinds of random diffusivity models whose difference is embodied in whether the fluctuation-dissipation theorem is valid. Based on the two random diffusivity models, we derive the Fokker-Planck equations with an arbitrary external force, and analyse various observables in the case with a constant force, including the Einstein relation, the moments, the kurtosis, and the asymptotic behaviors of the probability density function of particle's displacement at different time scales.
Both the theoretical results and numerical simulations of these observables show significant difference between the two kinds of random diffusivity models, which implies the important role of the fluctuation-dissipation theorem in random diffusivity systems.

\end{abstract}

\maketitle

\section{Introduction}\label{Sec1}
It is ubiquitous to find that particles diffuse under some kind of external force fields in the natural world. Under the effect of external forces, the motion of particles shows many kinds of anomalous diffusion phenomena in complex systems \cite{BouchaudGeorges:1990,MetzlerKlafter:2000,MagdziarzWeronKlafter:2008,EuleFriedrich:2009,CairoliBaule:2015,FedotovKorabel:2015}.
Particularly, the particles might undergo a biased random walk with a nonzero mean of displacement. The corresponding ensemble-averaged mean-squared displacement (MSD) is defined as
\begin{equation}\label{1}
\langle \Delta x^2(t)\rangle=\langle[x(t)-\langle x(t)\rangle]^2\rangle \,{\propto}\,t^\beta  \quad  (\beta\neq1),
\end{equation}
where normal Brownian motion belongs to $\beta=1$, and anomalous diffusion is characterized by the nonlinear evolution in time with $\beta\neq1$.

In addition to the normal diffusion of Brownian motion, the probability density function (PDF) of its displacement is Gaussian-shaped \cite{VanKampen:1992,CoffeyKalmykovWaldron:2004}
\begin{equation}\label{Gaussian}
  G(x,t|D)=\frac{1}{\sqrt{4\pi Dt}}\exp\left(-\frac{x^2}{4Dt}\right)
\end{equation}
for a given diffusivity $D$. In contrast to the Gaussian-shaped PDF, a new class of normal diffusion process has recently been observed with a non-Gaussian PDF, which is thus named as Brownian yet non-Gaussian process. This phenomenon has been found in a large range of complex systems, including polystyrene beads diffusing on the surface of lipid tubes \cite{WangAnthonyBaeGranick:2009} or in networks \cite{WangAnthonyBaeGranick:2009,ToyotaHeadSchmidtMizuno:2011,SilvaStuhrmannBetzKoenderink:2014}, as well as the diffusion of tracer molecules on polymer thin films \cite{Bhattacharya-etal:2013} and in simulations of two-dimensional discs \cite{KimKimSung:2013}. Instead of the Gaussian shape, the PDF of the Brownian yet non-Gaussian process is characterized by exponential distribution
\begin{equation}\label{Exponential}
  p(x,t)= \frac{1}{2\sqrt{D_0t}}\exp\left(-\frac{|x|}{\sqrt{D_0t}}\right)
\end{equation}
with $D_0$ being the effective diffusivity.

The interesting phenomenon of the non-Gaussian feature can be interpreted by the superstatistical approach of assuming the diffusivity $D$ in Eq. \eqref{Gaussian} being a random variable \cite{Beck:2001,BeckCohen:2003,Beck:2006}.
More precisely, each particle undergoes a normal Brownian motion with its own diffusivity which does not change considerably in a short time. The diffusivity $D$ of each particle obeys the exponential distribution
$\pi(D)=\exp(-D/D_0)/D_0$, and the randomness of diffusivity results from a spatially inhomogeneous environment. Averaging the Gaussian distribution in Eq. \eqref{Gaussian} over the diffusivity with the exponential distribution $\pi(D)$ yields \cite{WangKuoBaeGranick:2012,HapcaCrawfordYoung:2009}
\begin{equation}\label{PDF-x}
\begin{split}
    p(x,t)&=\int_0^\infty \pi(D)G(x,t|D)dD  \\
    &=\frac{1}{\sqrt{4D_0t}}\exp\left(-\frac{|x|}{\sqrt{D_0t}}\right).
\end{split}
\end{equation}
Besides the superstatistical approach, the exponential tail is found to be universal for short-time dynamics of the continuous-time random walk by using large deviation theory  \cite{BarkaiBurov:2020,WangBarkaiBurov:2020}.

Furthermore, the phenomenon observed in experiments also shows that the PDF undergoes a crossover from exponential distribution to Gaussian distribution \cite{WangAnthonyBaeGranick:2009,WangKuoBaeGranick:2012}. This crossover cannot reappear in the approach of the superstatistical dynamics. To interpret the phenomenon of such a crossover in the PDF of the Brownian yet non-Gaussian process, Chubynsky and Slater proposed a diffusing diffusivity model, in which the diffusion coefficient of the tracer particle evolves in time like the coordinate of a Brownian particle in a gravitational field \cite{ChubynskySlater:2014}. Chechkin {\it et al.} established a minimal model under the framework of Langevin equation with the diffusivity being the square of an Ornstein-Uhlenbeck process \cite{ChechkinSenoMetzlerSokolov:2017}. Due to the widespread applications of random diffusivity when describing the particle's motion in complex environments, the researches on systems with random parameters have been extended to many physical models, including underdamped Langevin equation \cite{SlezakMetzlerMagdziarz:2018,Vitali.etal:2018,ChenWang:2021}, generalized grey Brownian motion \cite{SposiniChechkinSenoPagniniMetzler:2018} and fractional Brownian motion \cite{JainSebastian:2018,MackalaMagdziarz:2019,Wang-etal:2020,Wang-etal:2020-2}, together with some discussions on ergodic property of random diffusivity systems \cite{CherstvyMetzler:2016,WangChen:2021,WangChen:2022}.

Our aim here is to consider the effect of an external force field on the Brownian yet non-Gaussian processes. Since it is convenient to describe a motion under an external force or an environment with fluctuation in a Langevin equation, we will investigate the effect of a force on the
minimal Langevin model with diffusing diffusivity proposed in Ref. \cite{ChechkinSenoMetzlerSokolov:2017}, where a Brownian particle with a random diffusivity $D(t)$ is described by
\begin{equation}\label{model0}
  \frac{d}{d t}x(t)=\sqrt{2D(t)}\xi(t).
\end{equation}
Here, $\xi(t)$ is the Gaussian white noise with mean zero and correlation function $\langle\xi(t_1)\xi(t_2)\rangle=\delta(t_1-t_2)$, and $D(t)$ is the square of an Ornstein-Uhlenbeck process to guarantee its positivity and randomness.

When considering the response of such a random diffusivity model to an external disturbance or the internal fluctuation of the system, we need to pay attention to whether the fluctuation-dissipation theorem (FDT) is valid or not in this system. The FDT plays a fundamental role in the statistical mechanics of nonequilibrium states and of irreversible processes \cite{Kubo:1966,MarconiPuglisiRondoniVulpiani:2008}. For this reason, two kinds of random diffusivity models, one satisfies FDT and one not, are considered, and their difference is also a main concerned object in this paper.

In addition to the FDT, Brownian motion also has a good property about Einstein relation which connects the fluctuation of an ensemble of particles with their mobility under a constant force $F$ by an equality \cite{Kubo:1966,MetzlerKlafter:2000}
\begin{equation}\label{ER}
  \langle x_F(t)\rangle=\frac{\langle x_0^2(t)\rangle}{2k_B\mathcal{T}}F.
\end{equation}
Here, $k_B$ is the Boltzmann constant, $\mathcal{T}$ is the absolute temperature of a heat bath, $x_F(t)$ and $x_0(t)$ denote the particle positions with and without the constant force $F$, respectively. Furthermore, the Einstein relation has been found to be valid for both normal and anomalous processes close to equilibrium in the limit $F\rightarrow0$, which can be derived from linear response theory \cite{BarkaiFleurov:1998,BenichouOshanin:2002,ShemerBarkai:2009,FroembergBarkai:2013-3}.
It will be interesting to find whether the Einstein relation holds or not in random diffusivity models.

In this paper, taking the two kinds of random diffusivity models satisfying the FDT or not as the main object, we first derive the Fokker-Planck equation of the PDF of particle's displacement for the two models under an arbitrary external force $F(x)$, and then make some specific analyses on the two models under a constant force $F$. The concerned observables mainly include the Einstein relation, the moments, the kurtosis of PDF, and the asymptotic behaviors of PDF.

The structure of this paper is as follows. In Sec. \ref{Sec2}, the two kinds of random diffusivity models are introduced. For arbitrary external force, the Fokker-Planck equations corresponding to the two models are derived in Sec. \ref{Sec3}. The detailed discussions on the observables for two models under a constant force are given in Secs. \ref{Sec4} and \ref{Sec5}, respectively. In Sec. \ref{Sec6}, we present the simulation results to verify the theoretical analyses on the observables for the case with constant force, and make a detailed comparison between the two models. Some discussions and summaries are provided in Sec. \ref{Sec5}. For convenience, we put some mathematical details in Appendix.

\section{Two random diffusivity models}\label{Sec2}
Since the FDT plays an important role on the diffusion behavior of a Langevin system, the difference between the two models concerned here is embodied in whether the FDT is valid or not.
Based on the random diffusivity model in Eq. \eqref{model0} characterizing the motion of a free particle, two kinds of models under an external force $F(x)$ can be written as
\begin{equation}\label{model-FDT}
  \frac{d}{d t}x(t)=\sqrt{2k_B\mathcal{T}D(t)}\xi(t)+D(t)F(x),
\end{equation}
and
\begin{equation}\label{model-NFDT}
  \frac{d}{d t}x(t)=\sqrt{2k_B\mathcal{T}D(t)}\xi(t)+F(x),
\end{equation}
respectively. The FDT is satisfied in Eq. \eqref{model-FDT}, which can be verified by dividing $D(t)$ on both sides, i.e.,
\begin{equation}\label{FDT-Eq}
  \frac{1}{D(t)}\frac{d}{d t}x(t)=\sqrt{\frac{2k_B\mathcal{T}}{D(t)}}\xi(t)+F(x).
\end{equation}
It can be seen that the dissipation memory kernel and correlation function of noise satisfy the relation \cite{Kubo:1966,KuboTodaHashitsume:1985,Zwanzig:2001,WangChenDeng:2019}
\begin{equation}\label{FDT-Relation}
  2k_B\mathcal{T}K(t_1-t_2)=\langle R(t_1)R(t_2)\rangle,
\end{equation}
where $K(t_1-t_2)=\delta(t_1-t_2)/D(t)$ is the dissipation memory kernel and $R(t)=\sqrt{\frac{2k_B\mathcal{T}}{D(t)}}\xi(t)$ is the internal noise in Eq. \eqref{FDT-Eq}. The FDT describes the phenomenon that the friction force and the random driving force come from the same origin and thus are closely related through Eq. \eqref{FDT-Relation}. For the Langevin system with a diffusing diffusivity $D(t)$ describing a spatially inhomogeneous environment, the FDT is still valid for each realization of $D(t)$.

Generalizing the idea in Refs. \cite{ChubynskySlater:2014,ChechkinSenoMetzlerSokolov:2017}, we use a generic overdampered Langevin equation to describe the diffusing diffusivity $D(t)$, i.e.,
\begin{equation}\label{model-DD}
\begin{split}
D(t)&=y^2(t),\\
\frac{d}{d t}y(t)&=f(y,t)+g(y,t) \eta(t),
\end{split}
\end{equation}
where the first equation is to guarantee the non-negativity of diffusivity $D(t)$, the second equation gives the evolution of auxiliary variable $y(t)$ with arbitrary functions $f(y,t)$ and $g(y,t)$ representing the external force and multiplicative noise on process $y(t)$. In addition, the noise $\eta(t)$ is also a Gaussion white noise with correlation function $\langle \eta(t_1)\eta(t_2)\rangle=\delta(t_1-t_2)$, similar to $\xi(t)$ but independent of $\xi(t)$. A special case that $f(y,t)=-y$ and $g(y,t)\equiv1$ yields the Ornstein-Uhlenbeck process $y(t)$ discussed in Ref. \cite{ChechkinSenoMetzlerSokolov:2017}.
Here, the arbitrary functions $f(y,t)$ and $g(y,t)$ in an overdamped Langevin equation result in a large range of diffusion processes beyond the Ornstein-Uhlenbeck process, including those reaching a steady state or not at long time limit, which is determined by the competitive roles between $f(y,t)$ and $g(y,t)$ \cite{WangDengChen:2019}. Many theoretical foundations have been established in the discussions on the ergodic properties and Feynman-Kac equations of the general overdamped Langevin equation \cite{WangDengChen:2019,WangChenDeng:2018,CairoliBaule:2017}.

\section{Fokker-Planck equations}\label{Sec3}
The Fokker-Planck equation governs the PDF $p(x,t)$ of finding the particle at position $x$ at time $t$, which describes the particle's stochastic motion in a macroscopic way. Compared with the Fokker-Planck equations containing integer derivatives for Brownian motion with or without an external force, those contain the fractional derivatives for many kinds of anomalous diffusion processes \cite{MetzlerKlafter:00,FriedrichJenkoBauleEule:2006,TurgemanCarmiBarkai:2009,KosztolowiczDutkiewicz:2021}. The Fokker-Planck equation for the random diffusivity model in Eq. \eqref{model0} have been derived in Ref. \cite{ChechkinSenoMetzlerSokolov:2017}.
Here we extend the model to the one containing an arbitrary external force and derive the corresponding Fokker-Planck equation.
Since the Langevin system includes three variables (the concerned process $x(t)$, diffusing diffusivity $D(t)$ and the auxiliary variable $y(t)$), and $D(t)$ depends on $y(t)$ explicitly as $D(t)=y^2(t)$, the bivariate PDF $p(x,y,t)$ is the underlying variable in the Fokker-Planck equation.
For convenience, we take $k_B\mathcal{T}=1$ in Eqs. \eqref{model-FDT} and \eqref{model-NFDT}, and take a space-dependent force $F(x)$. It should be noted that the results in this section are also valid for the case with time-dependent external force $F(x,t)$. The corresponding derivations can be obtained directly by replacing $F(x(s))$ with $F(x(s),s)$ in Eq. \eqref{split} and replacing $F(x(t))$ with $F(x(t),t)$ in Eq. \eqref{incre-x-y}.

Let us drive the Fokker-Planck equation corresponding to Eq. \eqref{model-FDT} firstly. Due to the FDT, the subordination method proposed in Ref. \cite{ChechkinSenoMetzlerSokolov:2017} for free particles can be applied here, i.e., rewriting the concerned process $x(t)$ as a compound process $x(t):=x(s(t))$ and splitting Eq. \eqref{model-FDT} into a Langevin system in subordinated form
\begin{equation}\label{split}
\begin{split}
\frac{d}{d s}x(s)&=\sqrt{2}\xi(s)+F(x(s)),\\
\frac{d}{d t}s(t)&=D(t),
\end{split}
\end{equation}
with the proof of the equivalence between them presented in Appendix \ref{App1}. The subordination method has been commonly used in Langevin system to describe subdiffusion \cite{Fogedby:1994,MetzlerKlafter:2000-2} or superdiffusion \cite{FriedrichJenkoBauleEule:2006,EuleZaburdaevFriedrichGeisel:2012,WangChenDeng:2019}.

The PDF $G(x,s)$ of process $x(s)$ in the first equation of Eq. \eqref{split} satisfies the classical Fokker-Planck equation \cite{Risken:1989,CoffeyKalmykovWaldron:2004}
\begin{equation}\label{Gxs}
\frac{\partial}{\partial s}G(x,s)=\left(-\frac{\partial}{\partial x}F(x)+\frac{\partial^2}{\partial x^2}\right)G(x,s).
\end{equation}
Combining the latter equation in Eq. \eqref{split}, we find $s(t)=\int_0^ty^2(t')dt'$. Therefore, $s(t)$ can be regarded as a functional of process $y(t)$, and the joint PDF $Q(s,y,t)$ satisfies the Feynman-Kac equation \cite{WangChenDeng:2018,CairoliBaule:2017,TurgemanCarmiBarkai:2009}
\begin{equation}\label{Qst}
\begin{split}
\frac{\partial}{\partial t}Q(s,y,t)&=\left(-\frac{\partial}{\partial y}f(y,t)+\frac{1}{2}\frac{\partial^2}{\partial y^2}g^2(y,t)\right)Q(s,y,t)\\
&~~~~-y^2\frac{\partial}{\partial s}Q(s,y,t).
\end{split}
\end{equation}
Since the two equations in Eq. \eqref{split} evolve independently, it holds that
\begin{equation}
\begin{split}
p(x,y,t)=\int_0^\infty G(x,s)Q(s,y,t)ds.
\end{split}
\end{equation}
Then combining the equations satisfied by $G(x,t)$ and $Q(s,y,t)$ in Eqs. \eqref{Gxs} and \eqref{Qst}, we obtain
\begin{equation}\label{FKE1}
\begin{split}
\frac{\partial}{\partial t}p(x,y,t)=&\int_0^\infty G(x,s)\frac{\partial}{\partial t}Q(s,y,t)ds\\
=&\left(-\frac{\partial}{\partial y}f(y,t)+\frac{1}{2}\frac{\partial^2}{\partial y^2}g^2(y,t)\right)p(x,y,t)\\
&-y^2\int_0^\infty G(x,s)\frac{\partial}{\partial s}Q(s,y,t) ds\\
=&\left(-\frac{\partial}{\partial y}f(y,t)+\frac{1}{2}\frac{\partial^2}{\partial y^2}g^2(y,t)\right)p(x,y,t)\\
&+y^2\left(-\frac{\partial}{\partial x}F(x)+\frac{\partial^2}{\partial x^2}\right)p(x,y,t),
\end{split}
\end{equation}
where the integration by parts has been used in the last equality and the corresponding boundary terms vanish.

For another model violating the FDT in Eq. \eqref{model-NFDT}, it cannot be split into two independent equations as Eq. \eqref{split}, and the subordination method is not applicable for this case. Instead, we adopt a universal Fourier transform method, which has been successfully used in deriving Fokker-Planck equation and Feynman-Kac equation \cite{DenisovHorsthemkeHanggi:2009,WangChenDeng:2018}.
Since the bivariate PDF $p(x,y,t)$ can be written as
$p(x,y,t)=\langle \delta(x-x(t))\delta(y-y(t))\rangle$, its Fourier transform ($x\rightarrow k_1, y\rightarrow k_2$) is
\begin{equation}
\begin{split}
  \tilde{p}(k_1,k_2,t)=& \int_{-\infty}^\infty\int_{-\infty}^\infty e^{-ik_1x-ik_2y}p(x,y,t)dxdy  \\
  =& \langle e^{-ik_1x(t)}e^{-ik_2y(t)}\rangle.
\end{split}
\end{equation}
The key point of this method is to derive the increment of $\tilde{p}(k_1,k_2,t)$ of order $\mathcal{O}(\tau)$ within a time interval $[t,t+\tau]$ when $\tau\rightarrow0$. Based on Eq. \eqref{model-NFDT} and the second equation of Eq. \eqref{model-DD}, we get the increments of $x(t)$ and $y(t)$ by omitting the higher order terms:
\begin{equation}\label{incre-x-y}
\begin{split}
x(t+\tau)-x(t)&\simeq\sqrt{2D(t)}\delta B_1(t)+F(x(t))\tau, \\[3pt]
y(t+\tau)-y(t)&\simeq f(y(t),t)\tau+g(y(t),t)\delta B_2(t),
\end{split}
\end{equation}
where $\delta B_i(t)=B_i(t+\tau)-B_i(t)$ is the increment of Brownian motion, $B_1(t)$ and $B_2(t)$ are independent from each other. By use of Eq. \eqref{incre-x-y}, the increment of $\tilde{p}(k_1,k_2,t)$ as $\delta \tilde{p}(k_1,k_2,t):=\tilde{p}(k_1,k_2,t+\tau)-\tilde{p}(k_1,k_2,t)$ can be evaluated as \\
\begin{widetext}
\begin{equation}\label{increment}
\begin{split}
\delta \tilde{p}(k_1,k_2,t)&=\langle e^{-ik_1x(t+\tau)}e^{-ik_2y(t+\tau)}\rangle-\langle e^{-ik_1x(t)}e^{-ik_2y(t)}\rangle\\
&\simeq \langle e^{-ik_1x(t)}e^{-ik_2y(t)}(e^{-ik_1(\sqrt{2D(t)}\delta B_1(t)+F(x(t))\tau)}e^{-ik_2(f(y(t),t)\tau+g(y(t),t)\delta B_2(t))}-1)\rangle\\
&\simeq \langle e^{-ik_1x(t)}e^{-ik_2y(t)}(-k_1^2D(t)\tau-ik_1F(x(t))\tau-ik_2f(y(t),t)\tau-\frac{1}{2}k_2^2g^2(y(t),t)\tau)\rangle,
\end{split}
\end{equation}
where we perform the ensemble average on $\delta B_1(t)$ and $\delta B_2(t)$ in the last line. More precisely, Eq. \eqref{incre-x-y} implies that both $x(t)$, $y(t)$ and $D(t)$ only depend on the increments $B_i$ of Brownian motion before time $t$, and thus they are independent from the increment $\delta B_i(t)$. We deal with the last two exponential functions in the second line by Taylor's series and only retain the terms of order $\mathcal{O}(\tau)$ as the last line shows.
Then dividing Eq. \eqref{increment} by $\tau$ on both sides, and taking the limit $\tau\rightarrow 0$, one arrives at
\begin{equation}
\begin{split}
\frac{\partial}{\partial t}\tilde{p}(k_1,k_2,t)=&-k^2_1\langle D(t)e^{-ik_1x(t)}e^{-ik_2y(t)}\rangle-ik_1\langle F(x(t))e^{-ik_1x(t)}e^{-ik_2y(t)}\rangle\\
&-ik_2\langle f(y(t),t)e^{-ik_1x(t)}e^{-ik_2y(t)}\rangle-\frac{1}{2}k^2_2\langle g^2(y(t),t)e^{-ik_1x(t)}e^{-ik_2y(t)}\rangle.
\end{split}
\end{equation}
\end{widetext}
Using the relation $D(t)=y^2(t)$ on the first term on the right-hand side, and performing inverse Fourier transform, we obtain the Fokker-Planck equation for the bivariate PDF $p(x,y,t)$ as
\begin{equation}\label{FKE2}
\begin{split}
\frac{\partial}{\partial t}p(x,y,t)&=\left(-\frac{\partial}{\partial x}F(x)+y^2\frac{\partial^2}{\partial x^2}\right)p(x,y,t)\\
&~~~+\left(-\frac{\partial}{\partial y}f(y,t)+\frac{1}{2}\frac{\partial^2}{\partial y^2}g^2(y,t)\right)p(x,y,t).
\end{split}
\end{equation}

Comparing the Fokker-Planck equations \eqref{FKE1} and \eqref{FKE2} for two different models in Eqs. \eqref{model-FDT} and \eqref{model-NFDT}, respectively, we find the main difference is embodied at the term containing external force $F(x)$. The former is $y^2F(x)$, i.e., $D(t)F(x)$ due to $D(t)=y^2(t)$ while the latter is $F(x)$. This difference is consistent to the discrepancy between the original models, i.e., $D(t)F(x)$ versus $F(x)$ in Eqs. \eqref{model-FDT} and \eqref{model-NFDT}. Actually, the Fokker-Planck equations \eqref{FKE1} can also be derived with the method of Fourier transform as Eq. \eqref{FKE2} by replacing $F(x)$ with $D(t)F(x)$ in the procedure.

Although the procedure of deriving the two Fokker-Planck equations looks a little complicated, the final form of Fokker-Planck equations can be understood in a simple way.
With a given $D(t)$, the corresponding Fokker-Planck equations governing the PDF $p(x,t)$ of displacement are
\begin{equation}\label{FK1}
\begin{split}
\frac{\partial}{\partial t}p(x,t)=D(t)\left[-\frac{\partial}{\partial x}F(x)+\frac{\partial^2}{\partial x^2}\right]p(x,t),
\end{split}
\end{equation}
and
\begin{equation}\label{FK2}
\begin{split}
\frac{\partial}{\partial t}p(x,t)=\left[-\frac{\partial}{\partial x}F(x)+D(t)\frac{\partial^2}{\partial x^2}\right]p(x,t)
\end{split}
\end{equation}
for Eqs. \eqref{model-FDT} and \eqref{model-NFDT}, respectively.
Then taking Eq. \eqref{FKE2} as an example, the terms on right-hand side can be divided into two parts. The first two terms are the ones in Fokker-Planck equation \eqref{FK2} by replacing $D(t)$ with $y^2$, while the last two terms come from the Fokker-Planck equation governing the PDF $p(y,t)$.
Albeit $D(t)$ is a diffusion process here, when we derive the Fokker-Planck equation governing the bivariate PDF $p(x,y,t)$, the role of $D(t)$ at the Fokker-Planck equation acts similarly to a deterministic function.

\section{Constant force field in Eq. (\ref{model-NFDT})}\label{Sec4}
For a comparison with the force-free case of Brownian yet non-Gaussian diffusion in Ref. \cite{ChechkinSenoMetzlerSokolov:2017}, we take $y(t)$ to be the Ornstein-Uhlenbeck process in the following discussions.
Let us first focus on the case that a constant force $F$ acts on the model \eqref{model-NFDT} where the FDT is broken. In this case, the Langevin system is written as
\begin{equation}\label{model_not_cons}
\begin{split}
\frac{d}{d t}x(t)&=\sqrt{2D(t)}\xi(t)+F,\\
D(t)&=y^2(t),\\
\frac{d}{d t}y(t)&=-y(t)+\eta(t).
\end{split}
\end{equation}
Based on the first equation, the process $x(t)$ can be written as
\begin{equation}\label{x-xf}
  x(t)=x_0(t)+Ft,
\end{equation}
where $x_0(t)$ denotes the trajectory of a free particle satisfying $dx_0(t)/dt=\sqrt{2D(t)}\xi(t)$ \cite{ChechkinSenoMetzlerSokolov:2017}. By the relation in Eq. \eqref{x-xf}, one has
\begin{equation}\label{relation}
\langle\Delta x^n(t)\rangle:=\langle (x(t)-\langle x(t)\rangle)^n\rangle=\langle x^n_0(t)\rangle,
\end{equation}
where $\langle x(t)\rangle=Ft$, and $x_0(t)$ is unbiased due to the symmetry of $\xi(t)$. Therefore, the ensemble-averaged MSD is $\langle \Delta x^2(t)\rangle=\langle x^2_0(t)\rangle\simeq t$.
The constant force here does not change the diffusion behavior and behaves as a decoupled force, which implies that model \eqref{model_not_cons} is Galilei invariant \cite{MetzlerKlafter:2000,CairoliKlagesBaule:2018,ChenWangDeng:2019-2}.
In addition, the drift $Ft$ dominates the diffusion process, and it holds that
\begin{equation}\label{moments1}
\langle x^n(t)\rangle\simeq F^nt^n.
\end{equation}
The relation between the first moment for the case with a constant force and the second moment of a free particle is
\begin{equation}
  \langle x(t)\rangle \simeq F\langle x^2_0(t)\rangle,
\end{equation}
which does not satisfy the Einstein relation in Eq. \eqref{ER}. This also relates to the violation of the FDT in Eq. \eqref{model-NFDT}.

Based on the moments, we can calculate the kurtosis to evaluate the deviation of the shape of a PDF from Gaussian distribution. The kurtosis of a one-dimensional Gaussian process is equal to $3$. Now for a biased process, the kurtosis is defined as
\begin{equation}\label{kurtosis}
\begin{split}
K=\frac{\langle \Delta x^4(t)\rangle}{\langle \Delta x^2(t)\rangle^2}.
\end{split}
\end{equation}
By use of Eq. \eqref{relation}, the kurtosis of the random diffusivity process under a constant force is
\begin{equation}\label{K1}
\begin{split}
K=\frac{\langle x^4_0(t)\rangle}{\langle x^2_0(t)\rangle^2}\simeq \left\{
\begin{array}{ll}
  9, &~ t\rightarrow 0, \\[5pt]
  3,  & ~t\rightarrow\infty,
\end{array}\right.
\end{split}
\end{equation}
consistent to the force-free case in Ref. \cite{ChechkinSenoMetzlerSokolov:2017}, where the PDF exhibits a crossover from exponential distribution to Gaussian distribution.

To be more delicate than the kurtosis, the explicit expression of the PDF $p(x,t)$ can be obtained through a translation of the PDF $p_0(x,t)$ of free particles to the positive direction with magnitude $Ft$, i.e.,
\begin{equation}\label{p1jianjin}
\begin{split}
p(x,t)&=p_0(x-Ft,t) \\[3pt]
&\simeq \left\{
\begin{array}{ll}
  \frac{1}{\pi t^{1/2}}K_0\left(\frac{x-Ft}{t^{1/2}}\right), &~ t\rightarrow 0,  \\[9pt]
  \frac{1}{(2\pi t)^{1/2}}\exp\left(-\frac{(x-Ft)^2}{2t}\right),  & ~t\rightarrow\infty,
\end{array}\right.
\end{split}
\end{equation}
where the expression of $p_0(x,t)$ is explicitly given in Eqs. (63) and (79) of Ref. \cite{ChechkinSenoMetzlerSokolov:2017} and $K_0$ is the Bessel function \cite{GradshteynRyzhikGeraniumsTseytlin:1980}.
In the short time limit, considering the asymptotics $K_0(z)\simeq \sqrt{\frac{\pi}{2z}}e^{-z}$ for $z\rightarrow\infty$, we have
\begin{equation}\label{PDF-short1}
\begin{split}
p(x,t)\simeq \frac{1}{\sqrt{2\pi|x-Ft|t^{1/2}}}\exp\left(-\frac{|x-Ft|}{t^{1/2}}\right),
\end{split}
\end{equation}
being an exponential distribution centered at $Ft$.

On the other hand, the short time asymptotics can be obtained from a superstatistical approach. For the time shorter than the diffusivity correlation time of the Ornstein-Uhlenbeck process, the diffusivity does not change considerably, and thus the initial condition in equilibrium of the Ornstein-Uhlenbeck process describes an ensemble of particles which diffuse with their own diffusion coefficient, resulting in a superstatistical result \cite{ChechkinSenoMetzlerSokolov:2017}.
In detail, the PDF $p_s(x,t)$ in superstatistical sense is given as the weighted average of a single Gaussian distribution $G(x,t|D)$ over the stationary distribution $p_D(D)$ of diffusivity $D$. The stationary distribution $p_D(D)$ can be obtained through the stationary distribution $f_{\textrm{st}}(y)=e^{-y^2}/\sqrt{\pi}$ of Ornstein-Uhlenbeck process in Eq. \eqref{model_not_cons}, i.e., \cite{ChechkinSenoMetzlerSokolov:2017}
\begin{equation}
\begin{split}
    p_D(D)=\int_{-\infty}^\infty f_{\textrm{st}}(y)\delta(D-y^2)dy
    =\frac{1}{\sqrt{\pi D}}e^{-D}.
\end{split}
\end{equation}
Then, it holds that
\begin{equation}
\begin{split}
p_s(x,t)&=\int_0^\infty p_D(D)G(x,t|D)dD  \\
&=\int_0^\infty \frac{1}{\sqrt{\pi D}}e^{-D} \cdot\frac{1}{\sqrt{4\pi Dt}}e^{-\frac{(x-Ft)^2}{4Dt}}dD \\
&=\frac{1}{\pi t^{1/2}}K_0\left(\frac{x-Ft}{t^{1/2}}\right),
\end{split}
\end{equation}
which is consistent to the short time asymptotics in Eq. \eqref{p1jianjin}.

\section{Constant force field in Eq. (\ref{model-FDT})}\label{Sec5}

The case that the constant force affects the diffusing diffusivity model \eqref{model-FDT} satisfying the FDT is
\begin{equation}\label{model_cons}
\begin{split}
\frac{d}{d t}x(t)&=\sqrt{2D(t)}\xi(t)+D(t)F,\\
D(t)&=y^2(t),\\
\frac{d}{d t}y(t)&=-y(t)+\eta(t).
\end{split}
\end{equation}
Similar to the way of deriving Fokker-Planck equation in Eq. \eqref{split}, it also brings convenience to rewrite the first equation of Eq. \eqref{model_cons} into a Langevin equation in the subordinated form, i.e.,
\begin{equation}\label{model_cons2}
\begin{split}
\frac{d}{d s}x(s)&=\sqrt{2}\xi(s)+F,\\
\frac{d }{d t}s(t)&=D(t),
\end{split}
\end{equation}
where the displacement is denoted as a compound process $x(t):=x(s(t))$.
Due to the independence between the two equations in Eq. \eqref{model_cons2}, it holds that
\begin{equation}\label{split-relation2}
\begin{split}
p(x,t)=\int_0^\infty G(x,s)O(s,t)ds,
\end{split}
\end{equation}
where $G(x,s)$ is the PDF of finding a Brownian particle under a constant force at position $x$ at time $s$, and $O(s,t)$ is the PDF of finding process $s(t)$ taking the value $s$ at time $t$. Therefore, $G(x,s)$ is a Gaussian distribution centered at $Fs$, i.e., $G(x,s)=\frac{1}{\sqrt{4\pi s}}e^{-\frac{(x-Fs)^2}{4s}}$ and $\tilde{G}(k,s)=e^{-ikFs-s k^2}$ in Fourier space ($x\rightarrow k$). Then we perform Fourier transform on Eq. \eqref{split-relation2} and obtain
\begin{equation}\label{pkt2}
\begin{split}
\tilde{p}(k,t)&=\int_0^\infty \tilde{G}(k,s)O(s,t)ds\\
&=\int_0^\infty e^{-(ikF+k^2)s} O(s,t)ds\\
&=\hat{O}(ikF+k^2,t),
\end{split}
\end{equation}
where $\hat{O}(ikF+k^2,t)$ denotes the Laplace transform $(s\rightarrow ikF+k^2)$ of the PDF $O(s,t)$. By use of the known result on the Laplace transform of $O(s,t)$ for the integrated square of the Ornstein-Uhlenbeck process \cite{Dankel:1991,ChechkinSenoMetzlerSokolov:2017}, we have
\begin{equation}\label{pkt-exact}
\begin{split}
    \tilde{p}(k,t)&=\exp\left(\frac{t}{2}\right)\left/\left[\frac{1}{2}\left(\sqrt{1+2\tilde{k}}+\frac{1}{\sqrt{1+2\tilde{k}}}\right)\right.\right.\\
&~~~\left.\times\textrm{sinh}\left(t\sqrt{1+2\tilde{k}}\right)
+\textrm{cosh}\left(t\sqrt{1+2\tilde{k}}\right)\right]^{\frac{1}{2}},
\end{split}
\end{equation}
where $\tilde{k}=ikF+k^2$. In order to satisfy the condition of Eq. \eqref{pkt-exact} proposed in Ref. \cite{Dankel:1991}, we assume that the initial position $y_0$ in Eqs. \eqref{model_not_cons} and \eqref{model_cons} obeys the equilibrium distribution of the Ornstein-Uhlenbeck process $y(t)$, i.e., a Gaussian distribution with mean zero and variance $1/2$:
\begin{equation}\label{EquilibriumDistribution}
  p_{\textrm{eq}}(y_0)=\frac{1}{\sqrt{\pi}}\exp(-y_0^2).
\end{equation}
This equilibrium distribution is also employed throughout all the simulations in Sec. \ref{Sec6}.
The expression of $\tilde{p}(k,t)$ in Eq. \eqref{pkt-exact} is exact for any time $t$, based on which we can evaluate the asymptotic moments and PDFs in $x$ space for short and long times.

For the moments, performing the Taylor expansion of exponential function in Eq. \eqref{pkt2} yields
\begin{equation}
\begin{split}
\tilde{p}(k,t)
&=\int_0^\infty e^{-\tilde{k}s} O(s,t)ds\\
&=1-\tilde{k}\langle s(t)\rangle+\frac{\tilde{k}^2}{2}\langle s^2(t)\rangle +\cdots.
\end{split}
\end{equation}
Then we use the formula $\langle x^n(t)\rangle=i^n\left.\frac{\partial^n}{\partial k^n}p(k,t)\right|_{k=0}$ and obtain the first four moments
\begin{equation}\label{Moments}
\begin{split}
\langle x(t)\rangle&=F\langle s(t)\rangle,\\
\langle x^2(t)\rangle&=2\langle s(t)\rangle+F^2\langle s^2(t)\rangle,\\
\langle x^3(t)\rangle&=6F\langle s^2(t)\rangle+F^3\langle s^3(t)\rangle,\\
\langle x^4(t)\rangle&=12\langle s^2(t)\rangle+12F^2\langle s^3(t)\rangle+F^4\langle s^4(t)\rangle.\\
\end{split}
\end{equation}
To obtain both short time and long time asymptotics, we need the accurate expressions of $\langle s^n(t)\rangle$, which are presented in Appendix \ref{App2}.
We find that for long times,
\begin{equation}\label{moments2}
  \langle x^n(t)\rangle\simeq \frac{F^n}{2^n}t^n.
\end{equation}
The relation between the first moment for the case with a constant force and the second moment for the force-free case is
\begin{equation}\label{ER-S}
  \langle x(t)\rangle \simeq \frac{F}{2}\langle x^2_0(t)\rangle,
\end{equation}
which satisfies the Einstein relation in Eq. \eqref{ER}.

Based on Eq. \eqref{Moments} and the accurate expression of $\langle s^n(t)\rangle$ in Appendix \ref{App2}, the MSD is equal to
\begin{equation}\label{M2}
\begin{split}
\langle\Delta x^2(t)\rangle&=\left(\frac{F^2}{2}+1\right)t+\frac{F^2}{4}(e^{-2t}-1)\\
&\simeq \left\{
\begin{array}{ll}
  t, &~t\rightarrow 0,  \\[5pt]
  \left(\frac{F^2}{2}+1\right)t,  & ~t\rightarrow \infty.
\end{array}\right.
\end{split}
\end{equation}
When $F=0$, it recovers to the constantly normal diffusion $\langle x^2(t)\rangle=t$.
Under the influence of a constant force, the particles still exhibit normal diffusion, but the effective diffusion coefficient increases from $1$ to $F^2/2+1$ as time goes.
Similar to the MSD in Eq. \eqref{M2}, the asymptotic expressions of fourth moment can be obtained from Eqs. \eqref{Moments} and Appendix \ref{App2}:
\begin{equation}\label{M4}
\begin{split}
\langle \Delta x^4(t)\rangle\simeq \left\{
\begin{array}{ll}
  9t^2, &~t\rightarrow 0,  \\[5pt]
  3\left(\frac{F^2}{2}+1\right)^2t^2,  & ~t\rightarrow \infty.
\end{array}\right.
\end{split}
\end{equation}
The constant force enhances the diffusion slightly since it only increases the diffusion coefficient without changing the diffusion behavior at long time limit.

Here we also evaluate the kurtosis to predict the shape of the PDF $p(x,t)$ for the case satisfying FDT. Considering the definition of kurtosis in Eq. \eqref{kurtosis}, and combining the moments in Eqs. \eqref{M2} and \eqref{M4}, we find
\begin{equation}\label{K2}
\begin{split}
K\simeq \left\{
\begin{array}{ll}
  9, &~ t\rightarrow 0, \\[5pt]
  3,  & ~t\rightarrow\infty.
\end{array}\right.
\end{split}
\end{equation}
Surprisingly, this result is consistent to the force-free case and the result in Eq. \eqref{K1}, which implies a possible crossover of PDF from exponential distribution to Gaussian distribution as the force-free case.

For the asymptotic expression of PDF $p(x,t)$, taking $t\rightarrow 0$ in Eq. \eqref{pkt-exact} yields
\begin{equation}\label{pkt-ST}
\begin{split}
\tilde{p}(k,t)\simeq t^{-\frac{1}{2}}\left(ikF+k^2+\frac{1}{t}\right)^{-\frac{1}{2}}.
\end{split}
\end{equation}
The normalization of the asymptotic PDF can be verified by $\tilde{p}(k=0,t)=1$. The inverse Fourier transform of $\tilde{p}(k,t)$ cannot be obtained easily. Since $t\rightarrow 0$, whenever $k\rightarrow0$ or $k\rightarrow\infty$, the imaginary part in the brackets of Eq. \eqref{pkt-ST} is much smaller than the real part, i.e., $kF\ll k^2+1/t$. Therefore, the constant force $F$ here only makes a slight biase on the original PDF. The expression of the biased PDF will be explicitly given through a superstatistical approach in the following. The asymptotic behavior at short time limit should be consistent to the corresponding superstatistical result.

In superstatistical approach, the effective PDF $p_s(x,t)$ is given as the weighted average of the conditional Gaussian distribution over the stationary distribution $p_D(D)$, i.e.,
\begin{equation}\label{short}
\begin{split}
p_s(x,t)&=\int_0^\infty p_D(D)G(x,t|D)dD  \\
&=\frac{1}{\sqrt{4\pi^2t}}e^{\frac{Fx}{2}}\int_0^\infty \frac{1}{D}e^{-D\left(1+\frac{F^2}{4}t\right)}e^{-\frac{x^2}{4Dt}}dD\\
&=\frac{1}{\pi\sqrt{t}}e^{\frac{Fx}{2}}K_0\left(\frac{\sqrt{4+F^2t}x}{2\sqrt{t}}\right),
\end{split}
\end{equation}
where  $G(x,t|D)=\frac{1}{\sqrt{4\pi Dt}}e^{-\frac{(x-FDt)^2}{4Dt}}$ has been used.
Then using the asymptotic behavior $K_0(z)\simeq \sqrt{\frac{\pi}{2z}}e^{-z}$ as $z\rightarrow\infty$, we arrive at
\begin{equation}\label{psxt}
\begin{split}
p_s(x,t)&\simeq \frac{1}{\sqrt{2\pi|x|t^{1/2}}}\frac{1}{\sqrt{(1+F^2t/4)^{1/2}}}  \\
&~~~~\times
\exp\left(\frac{Fx}{2}-\sqrt{1+F^2t/4}\frac{|x|}{t^{1/2}}\right).
\end{split}
\end{equation}
Corresponding to the short time aymptotics in Eq. \eqref{pkt-ST}, we take $t\ll 4/F^2$ in Eq. \eqref{psxt}, and obtain
\begin{equation}\label{psxt-ST}
\begin{split}
p_s(x,t)\simeq p_0(x,t) \exp\left(\frac{Fx}{2}\right),
\end{split}
\end{equation}
where
\begin{equation}\label{p0xt}
  p_0(x,t)=\frac{1}{\sqrt{2\pi|x|t^{1/2}}}\exp\left(-\frac{|x|}{t^{1/2}}\right)
\end{equation}
is the PDF of free particles in the superstatistical case. It can be seen that the constant force only adds a time-independent correction $e^{Fx/2}$ to the PDF of free particles at short time limit. Compared with the exponential part in $p_0(x,t)$, the exponential correction $e^{Fx/2}$ is negligible for short time since the exponential coefficient satisfies $F/2\ll 1/t^{1/2}$. This result is consistent to the previous kurtosis $K\simeq9$ in Eq. \eqref{K2} at short time limit and the analyses following Eq. \eqref{pkt-ST}.

On the other hand, the long time asymptotics $t\gg 4/F^2$ of $p_s(x,t)$ is
\begin{equation}\label{psxt-LT}
\begin{split}
p_s(x,t)\simeq p_0(x,t)C_F(x,t),
\end{split}
\end{equation}
where
\begin{equation}
  \begin{split}
    C_F(x,t)=\left\{
    \begin{array}{ll}
    \frac{1}{\sqrt{Ft^{1/2}/2}}\exp\left(\frac{x}{2t^{1/2}}\right),  &  x>0,  \\[5pt]
    \frac{1}{\sqrt{Ft^{1/2}/2}}\exp\left(Fx\right),  & x<0.
    \end{array}\right.
  \end{split}
\end{equation}
The constant force makes the PDF biased to the positive direction, i.e., decaying more slowly for $x>0$ but faster for $x<0$. Furthermore, the change in PDF at $x<0$ is more obvious than that at $x>0$. For long time limit, the exponential coefficient $F$ in $C_F(x,t)$ is much larger than $t^{-1/2}$ in $p_0(x,t)$, i.e., $F\gg 1/t^{1/2}$. So the dominating term of decaying when $x<0$ is $e^{Fx}$.

In contrast to the superstatistical results above, the real long time asymptotics of the Langevin system in Eq. \eqref{model_cons} can be found by taking $t\rightarrow \infty$ in Eq. \eqref{pkt-exact}. The asymptotic result is
\begin{equation}
\begin{split}
\tilde{p}(k,t)\simeq \frac{\sqrt{2}\exp\left(\frac{t}{2}(1-\sqrt{1+2\tilde{k}})\right)}
{\left[\frac{1}{2}\left(\sqrt{1+2\tilde{k}}+\frac{1}{\sqrt{1+2\tilde{k}}}\right)+1\right]^{1/2}}.
\end{split}
\end{equation}
Then we consider the large-$x$ behavior by taking $k\rightarrow 0$, and obtain
\begin{equation}
\begin{split}
\tilde{p}(k,t)\simeq \exp\left(-\frac{iFt}{2}k -\frac{(2+F^2)t}{4}k^2\right).
\end{split}
\end{equation}
With the inverse Fourier transform, the Gaussian distribution with mean $Ft/2$ and variance $(F^2/2+1)t$ is obtained:
\begin{equation}\label{long}
\begin{split}
p(x,t)\simeq \frac{1}{\sqrt{2\pi\left(\frac{F^2}{2}+1\right)t}}
\exp\left(-\frac{\left(x-\frac{F}{2}t\right)^2}{2\left(\frac{F^2}{2}+1\right)t}\right).
\end{split}
\end{equation}
This Gaussian shape is also consistent to the previous kurtosis $K\simeq3$ in Eq. \eqref{K2} at long time limit.

\section{Simulations}\label{Sec6}
\begin{figure}
  \centering
  \includegraphics[scale=0.5]{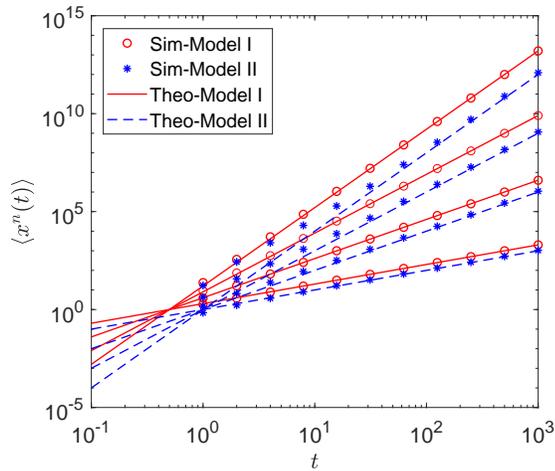}\\
  \caption{Moments $\langle x^n(t)\rangle$ with $n=1,2,3,4$. Model I (in red) and Model II (in blue) represent the Langevin systems in Eqs. \eqref{model_not_cons} and \eqref{model_cons}, respectively.
  The circle and star markers denote the simulation results, while the solid and dashed lines denote the theoretical results in Eqs. \eqref{moments1} and \eqref{moments2}, respectively. Based on Eqs. \eqref{moments1} and \eqref{moments2}, the two lines with the same $n$ are parallel for two models, i.e., $\langle x^n(t)\rangle_I=2^n\langle x^n(t)\rangle_{I\!I}$. Correspondingly, each two lines (or markers) from the bottom to the top represent the first, second, third, and fourth moments, respectively. Parameters: $T=10^3$, $F=2$, and $10^3$ samples are used for ensemble average.
}\label{fig1}
\end{figure}

\begin{figure}
  \centering
  \includegraphics[scale=0.5]{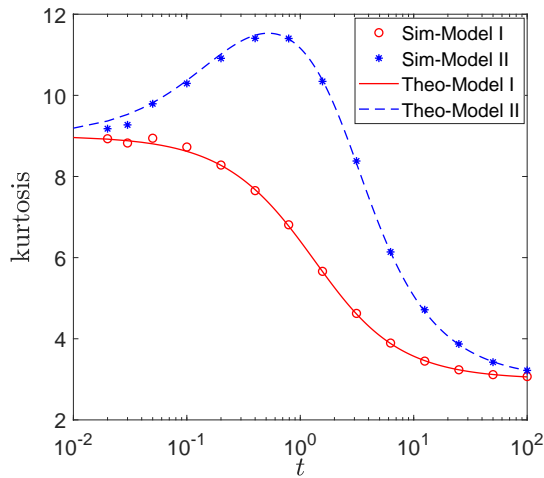}\\
  \caption{Kurtosis (defined in Eq. \eqref{kurtosis}) in Model I (in red) and Model II (in blue) which represent the Langevin systems in Eqs. \eqref{model_not_cons} and \eqref{model_cons}, respectively. For the two models, the circle and star markers denote the simulation results, while the solid and dashed lines denote the theoretical results in Eqs. \eqref{K1-exact} and \eqref{K2-exact}, respectively. The kurtosis in Eqs. \eqref{K1} and \eqref{K2} both have the same asymptotics as the force-free case.
In contrast to the monotone decreasing behavior of the kurtosis line of Model I, that of Model II has a maximum value around $t=0.5$. Parameters: $T=10^2$, $F=2$, and $10^6$ samples are used for ensemble average.
}\label{fig2}
\end{figure}

\begin{figure}
  \centering
  \includegraphics[scale=0.5]{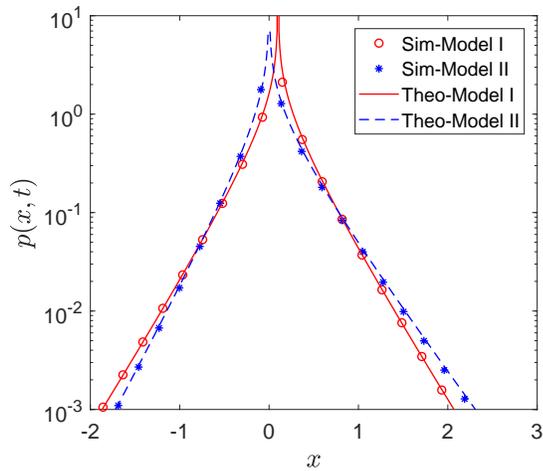}\\
  \caption{Short-time PDFs in Model I (in red) and Model II (in blue) which represent the Langevin systems in Eqs. \eqref{model_not_cons} and \eqref{model_cons}, respectively. For the two models, the circle and star markers denote the simulation results, while the solid and dashed lines denote the theoretical results in Eqs. \eqref{PDF-short1} and \eqref{psxt-ST}, respectively. The PDF of Model I is a symmetric exponential distribution with the center at $x=Ft$, while the PDF of Model II is an asymmetric skewed exponential distribution.
Parameters: $T=0.1$, $F=1$, and $10^7$ samples are used for ensemble average.}\label{fig3}
\end{figure}

\begin{figure}
  \centering
  \includegraphics[scale=0.5]{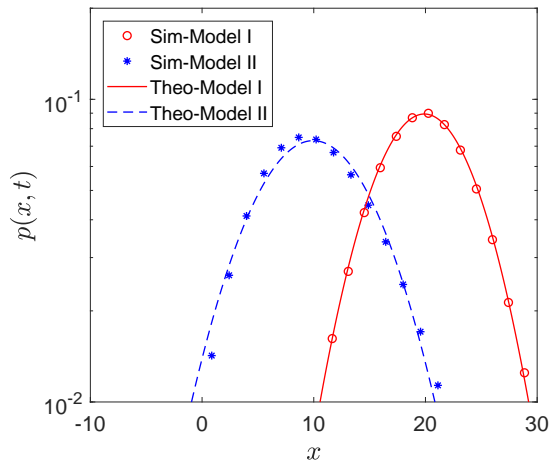}\\
  \caption{Long-time PDFs in Model I (in red) and Model II (in blue) which represent the Langevin systems in Eqs. \eqref{model_not_cons} and \eqref{model_cons}, respectively. For the two models, the circle and star markers denote the simulation results, while the solid and dashed lines denote the theoretical results in Eqs. \eqref{p1jianjin} and \eqref{long}, respectively. Both the PDFs of two models are Gaussian shapes. The PDF of Model I has the mean $Ft$ and the variance $t$, while the one of Model II has a smaller mean $Ft/2$ but a larger variance $(F^2/2+1)t$.
Parameters: $T=20$, $F=1$, and $10^7$ samples are used for ensemble average.}\label{fig4}
\end{figure}

In all our simulations, the initial position $y_0$ of the Langevin systems in Eqs. \eqref{model_not_cons} and \eqref{model_cons} is taken from the equilibrium distribution $N(0,1/\sqrt{2})$ in Eq. \eqref{EquilibriumDistribution}, and the two models in Eqs. \eqref{model_not_cons} and \eqref{model_cons} are recorded briefly as ``Model I'' and ``Model II'', respectively.
For a clear comparison between the two models, we put the simulation results of the same observable in one figure, with their moments in Fig. \ref{fig1}, kurtosis in Fig. \ref{fig2}, short-time PDFs in Fig. \ref{fig3}, and long-time PDFs in Fig. \ref{fig4}.

In Fig. \ref{fig1}, we simulate the first four moments $\langle x^n(t)\rangle$ of two models, which agree with the theoretical results very well. According to the theoretical results in Eqs. \eqref{moments1} and \eqref{moments2}, we find the moments of two models only differ by a constant multiplier, i.e.,
\begin{equation}
  \langle x^n(t)\rangle_I=2^n\langle x^n(t)\rangle_{I\!I}.
\end{equation}
As a result, the solid and dashed lines (or circle and star markers) in Fig. \ref{fig1} are parallel for the same $n$.

In Fig. \ref{fig2}, we simulate the kurtosis for two models. They have the same asymptotic results in Eqs. \eqref{K1} and \eqref{K2} with a crossover from $K=9$ at the beginning to $K=3$ at the infinity. In addition to the asymptotic results, the exact expressions of kurtosis can be obtained by use of the definition in Eq. \eqref{kurtosis} and the first four moments $\langle x^n(t)\rangle$ in Eqs. \eqref{relation} and \eqref{Moments}. For convenience, the exact expressions are presented in Appendix \ref{App3}, where the latter (Eq. \eqref{K2-exact}) recovers the former (Eq. \eqref{K1-exact}) when $F=0$. The kurtosis of Model I is the same as the force-free case \cite{ChechkinSenoMetzlerSokolov:2017} due to its Galilei invariant property.
In contrast to the monotone decreasing kurtosis from $9$ to $3$ in Model I, the kurtosis of Model II has a maximum value around $t=0.5$, which means that for short time, the PDF of Model II undergoes a significant deviation from the Gaussian distribution. The reason can be found from the asymptotic PDF at short time limit in Eq. \eqref{psxt-ST}. The additional term $e^{Fx/2}$ brings a biase to the original exponential distribution $p_0(x,t)$ in Eq. \eqref{p0xt}. At long time limit, the PDF converges to the Gaussian distribution in Eq. \eqref{long}, corresponding to the monotone decreasing kurtosis after $t=0.5$ in Model II.

The asymptotic PDFs of two models for short time are presented in Fig. \ref{fig3}. The corresponding theoretical results are given in Eqs. \eqref{PDF-short1} and \eqref{psxt-ST}, respectively. For Model I, the PDF is exactly a translation to the positive direction with the magnitude $x=Ft$ of the original PDF $p_0(x,t)$ for force-free case. In contrast to Model I, the PDF of Model II is asymmetric due to the term $e^{Fx/2}$ in Eq. \eqref{psxt-ST}. It can be found that the lines in a semi-log graph (Fig. \ref{fig3}) are not exactly straight. The slight deviation from straight lines comes from the power-law correction term $|x|^{-1/2}$ in $p_0(x,t)$ in Eq. \eqref{p0xt}.

The asymptotic PDFs of two models for long time are presented in Fig. \ref{fig4}. The corresponding theoretical results are given in Eqs. \eqref{p1jianjin} and \eqref{long}, respectively. Corresponding to the behavior of the kurtosis tending to $3$ in Fig. \ref{fig2}, the PDFs for two models both converge to the Gaussian distribution at long time limit. As the shape of PDFs in Fig. \ref{fig4} shows, the PDF of Model I has the mean $Ft$ and the variance $t$, while the one of Model II has a smaller mean $Ft/2$ but a larger variance $(F^2/2+1)t$. This feature comes from the fact that the constant force $F$ is multiplied by a stochastic process $D(t)$ which enhances the fluctuation, and that the mean of $D(t)$ at steady state is $1/2$ which weakens the effective drift by half.

\section{Conclusion}\label{Sec7}
Much attention has been taken to the scenarios of how external force (or constant force) influences a dynamic system with a power-law distributed waiting time \cite{BarkaiFleurov:1998,MetzlerKlafter:2000,FroembergBarkai:2013-3,ChenWangDeng:2019-2,ChenWangDeng:2019-3}.
This paper extends this issue to the random diffusivity model with a diffusing diffusivity $D(t)$, and explores how the diffusing diffusivity $D(t)$ acts in a system under an external force.
Considering the importance of the FDT in the statistical mechanics of nonequilibrium dynamics, we build two kinds of random diffusivity models with an external force based on whether the FDT satisfies or not.

The main studies on the two models can be divided into two parts: one derives the Fokker-Planck equation of random diffusivity models with arbitrary external force, and another one investigates in detail some common quantities by taking a specific constant force.
In the first part, the Fokker-Planck equations for the bivariate PDF $p(x,y,t)$ of
two random diffusivity models under an arbitrary external force field are derived in Eqs. \eqref{FKE1} and \eqref{FKE2}. Corresponding to the fact that the only difference between the original Langevin equations \eqref{model-FDT} and \eqref{model-NFDT} is $F(x)$ versus $D(t)F(x)$, the difference between the Fokker-Planck equations is only embodied at the external force term, $F(x)$ versus $y^2F(x)$. Although $D(t)$ is a diffusion process, the role of $D(t)$ at the expression of Fokker-Planck equations is similar to a deterministic function.
The structure of the derived Fokker-Planck equations has striking character. Due to the independence between the evolution of concerned process $x(t)$ and auxiliary process $y(t)$, the right-hand side of Fokker-Planck equations \eqref{FKE1} and \eqref{FKE2} can be divided into two parts, being the terms in the corresponding Fokker-Planck equation governing the PDF $p(x,t)$ and $p(y,t)$, respectively.

In the second part, we investigate the case with constant force field and the diffusivity $D(t)$ being
the square of Ornstein-Uhlenbeck process by studying the moments, Einstein relation, the kurtosis and the asymptotic behaviors of the PDF in detail.
For random diffusivity model in Eq. \eqref{model_not_cons} with the FDT broken, we establish the relation between the concerned process $x(t)$ under the effect of a constant force and the displacement $x_0(t)$ of a free particle by $x(t)=x_0(t) +Ft$. Thus we find this model is Galilei invariant, similar to the discussed anomalous processes \cite{MetzlerKlafter:2000,CairoliKlagesBaule:2018,ChenWangDeng:2019-2}. The diffusion behavior is not changed by the constant force. The mean value is $Ft$ and the Einstein relation is not valid in this model. Compared with the PDF of force-free case, the PDF is translated to the positive direction with a biase $Ft$, with the kurtosis and the asymptotic behaviors of PDF unchanged.

For the random diffusivity model in Eq. \eqref{model_cons} satisfying the FDT, the results are quite different from the force-free case. The theoretical derivations are based on the technique of splitting the first equation of Eq. \eqref{model_cons} into a Langevin equation in subordinated form. We find the mean value of displacement is $\langle x(t)\rangle=Ft/2$ in this case, satisfying the Einstein relation Eq. \eqref{ER-S}. Although the kurtosis has the same asymptotic behavior at $t\rightarrow0$ and $t\rightarrow\infty$, it is not monotone any more. It increases for short time and reaches the maximum around $t=0.5$ as Fig. \ref{fig2} shows.
For long time, the PDF surprisingly converges to a Gaussian distribution as the force-free case, while the PDF in short-time limit is biased due to a correction $e^{Fx/2}$ compared with the force-free case.

Many significant differences between the two models imply that the FDT also plays an important role in random diffusivity systems. Through detailed analyses on the kurtosis and the shape of PDF, the model satisfying the FDT shows many interesting dynamic behaviors due to the existence of random diffusivity $D(t)$. These results will bring benefits to the discussions on how anomalous diffusion particles response to the external force in more random diffusivity systems.

\section*{Acknowledgments}
This work was supported by the National Natural Science
Foundation of China under Grant No. 12105145, the Natural Science Foundation of Jiangsu Province under Grant No. BK20210325.

\appendix

\section{Equivalence between Eqs. \eqref{model-FDT} and \eqref{split}}\label{App1}
The main idea of proving the equivalence is to combine the two equations in Eq. \eqref{split} and to transform them into Eq. \eqref{model-FDT}. Noting that the diffusing diffusivity $D(t)$ is independent of the noise $\xi$, $D(t)$ can be regarded as a deterministic function and the ensemble average only acts on $\xi$ in the following.
Integrating the first equation in Eq. \eqref{split} yields
\begin{equation}\label{A1}
  x(s)=\sqrt{2}\int_0^s\xi(s')ds'+\int_0^sF(x(s'))ds',
\end{equation}
where we have assumed the initial condition $x(0)=0$.
Since the concerned process $x(t)$ has been written as a compound process $x(t):=x(s(t))$, $x(t)$ can be obtained by replacing $s$ with $s(t)$ in Eq. \eqref{A1}, i.e.,
\begin{equation}\label{A2}
  x(t)=\sqrt{2}\int_0^{s(t)}\xi(s')ds'+\int_0^{s(t)}F(x(s'))ds'.
\end{equation}
By using the second equation of Eq. \eqref{split} and performing the derivative over time $t$ on both sides of Eq. \eqref{A2}, one arrives at
\begin{equation}\label{A3}
  \frac{d}{dt}x(t)=\sqrt{2}D(t)\xi(s(t))+D(t)F(x(t)).
\end{equation}
Now the only difference between Eqs. \eqref{A3} and \eqref{model-FDT} is the first term on the right-hand side. It is sufficient to prove that they share the same correlation function since $\xi$ is white Gaussian noise. A formula about $\delta$-function
\begin{equation}
  \delta(h(x))=\sum_{i}\frac{\delta(x-x_i)}{|h'(x_i)|}
\end{equation}
will be used, where $x_i$ is the $i$th simple root of $h(x)=0$. Utilizing this formula and a truth that $s(t)$ is monotone increasing, we have
\begin{equation}
\begin{split}
    \langle\xi(s(t_1))\xi(s(t_2))\rangle
&=\delta(s(t_1)-s(t_2))  \\
&=\frac{1}{D(t_1)}\delta(t_1-t_2).
\end{split}
\end{equation}
Therefore, it can be found that both the correlation functions of the first term in Eqs. \eqref{A3} and \eqref{model-FDT} are
\begin{equation}
  2D(t_1)\delta(t_1-t_2).
\end{equation}

\section{Moments of process $s(t)$}\label{App2}
The moments of process $s(t)$ in Eq. \eqref{model_cons2} can be obtained from its PDF in Laplace space by use of the formula
\begin{equation}
  \langle s^n(t)\rangle=(-1)^n\left.\frac{\partial^n}{\partial \lambda^n}\hat{O}(\lambda,t)\right|_{\lambda=0},
\end{equation}
where $\hat{O}(\lambda,t)$ is the Laplace transform of $O(s,t)$, and \cite{Dankel:1991,ChechkinSenoMetzlerSokolov:2017}
\begin{equation}
\begin{split}
    \hat{O}(\lambda,t)&=\exp\left(\frac{t}{2}\right)\left/\left[\frac{1}{2}\left(\sqrt{1+2\lambda}+\frac{1}{\sqrt{1+2\lambda}}\right)\right.\right.\\
&~~~\left.\times\textrm{sinh}\left(t\sqrt{1+2\lambda}\right)
+\textrm{cosh}\left(t\sqrt{1+2\lambda}\right)\right]^{\frac{1}{2}}.
\end{split}
\end{equation}
\vspace{5mm}
With some tedious calculations, it holds that
\begin{equation}
\langle s(t)\rangle=\frac{t}{2},
\end{equation}
\begin{equation*}
\langle s^2(t)\rangle=\frac{1}{4}(e^{-2t}-1+2t+t^2),
\end{equation*}
\begin{equation*}
\langle s^3(t)\rangle=\frac{1}{8}\Big(3(4+5t)e^{-2t}-12+9t+6t^2+t^3\Big),
\end{equation*}
\begin{equation*}
\begin{split}
\langle s^4(t)\rangle&=\frac{1}{16}\Big(6(27+50t+25t^2)e^{-2t}+9e^{-4t} \\[2pt]
   &~~~~-171+60t+54t^2+12t^3+t^4\Big).
\end{split}
\end{equation*}

\section{Exact kurtosis}\label{App3}
The exact theoretical expressions of kurtosis for two models in Eqs. \eqref{model_not_cons} and \eqref{model_cons} are
\begin{equation}\label{K1-exact}
  K=\frac{3}{t^2}(-1+e^{-2t}+2t+t^2)
\end{equation}
for Model I, and
\begin{widetext}
\begin{equation}\label{K2-exact}
  \begin{split}
  K&=\left\{-3-18F^2-\frac{171}{16}F^4+\left(6+\frac{33}{2}F^2+\frac{27}{4}F^4\right)t
    +\left(3+3F^2+\frac{3}{4}F^4\right)t^2 \right.\\
&~~~~+ \left. \left[3+\left(18+\frac{39}{2}t\right)F^2
    +\left(\frac{81}{8}+\frac{63}{4}t+6t^2\right)F^4\right]e^{-2t}+\frac{9}{16}F^4e^{-4t}  \right\}
\left/ \left[\left(\frac{F^2}{2}+1\right)t+\frac{F^2}{4}(e^{-2t}-1)\right]^2                        \right.
\end{split}
\end{equation}
for Model II.
\end{widetext}

\section*{References}
\bibliography{ReferenceCW}

\begin{thebibliography}{59}%
\makeatletter
\providecommand \@ifxundefined [1]{%
 \@ifx{#1\undefined}
}%
\providecommand \@ifnum [1]{%
 \ifnum #1\expandafter \@firstoftwo
 \else \expandafter \@secondoftwo
 \fi
}%
\providecommand \@ifx [1]{%
 \ifx #1\expandafter \@firstoftwo
 \else \expandafter \@secondoftwo
 \fi
}%
\providecommand \natexlab [1]{#1}%
\providecommand \enquote  [1]{``#1''}%
\providecommand \bibnamefont  [1]{#1}%
\providecommand \bibfnamefont [1]{#1}%
\providecommand \citenamefont [1]{#1}%
\providecommand \href@noop [0]{\@secondoftwo}%
\providecommand \href [0]{\begingroup \@sanitize@url \@href}%
\providecommand \@href[1]{\@@startlink{#1}\@@href}%
\providecommand \@@href[1]{\endgroup#1\@@endlink}%
\providecommand \@sanitize@url [0]{\catcode `\\12\catcode `\$12\catcode
  `\&12\catcode `\#12\catcode `\^12\catcode `\_12\catcode `\%12\relax}%
\providecommand \@@startlink[1]{}%
\providecommand \@@endlink[0]{}%
\providecommand \url  [0]{\begingroup\@sanitize@url \@url }%
\providecommand \@url [1]{\endgroup\@href {#1}{\urlprefix }}%
\providecommand \urlprefix  [0]{URL }%
\providecommand \Eprint [0]{\href }%
\providecommand \doibase [0]{https://doi.org/}%
\providecommand \selectlanguage [0]{\@gobble}%
\providecommand \bibinfo  [0]{\@secondoftwo}%
\providecommand \bibfield  [0]{\@secondoftwo}%
\providecommand \translation [1]{[#1]}%
\providecommand \BibitemOpen [0]{}%
\providecommand \bibitemStop [0]{}%
\providecommand \bibitemNoStop [0]{.\EOS\space}%
\providecommand \EOS [0]{\spacefactor3000\relax}%
\providecommand \BibitemShut  [1]{\csname bibitem#1\endcsname}%
\let\auto@bib@innerbib\@empty
\bibitem [{\citenamefont {Bouchaud}\ and\ \citenamefont
  {Georges}(1990)}]{BouchaudGeorges:1990}%
  \BibitemOpen
  \bibfield  {author} {\bibinfo {author} {\bibfnamefont {J.-P.}\ \bibnamefont
  {Bouchaud}}\ and\ \bibinfo {author} {\bibfnamefont {A.}~\bibnamefont
  {Georges}},\ }\bibfield  {title} {\bibinfo {title} {Anomalous diffusion in
  disordered media: {S}tatistical mechanisms, models and physical
  applications},\ }\href@noop {} {\bibfield  {journal} {\bibinfo  {journal}
  {Phys. Rep.}\ }\textbf {\bibinfo {volume} {195}},\ \bibinfo {pages} {127}
  (\bibinfo {year} {1990})}\BibitemShut {NoStop}%
\bibitem [{\citenamefont {Metzler}\ and\ \citenamefont
  {Klafter}(2000{\natexlab{a}})}]{MetzlerKlafter:2000}%
  \BibitemOpen
  \bibfield  {author} {\bibinfo {author} {\bibfnamefont {R.}~\bibnamefont
  {Metzler}}\ and\ \bibinfo {author} {\bibfnamefont {J.}~\bibnamefont
  {Klafter}},\ }\bibfield  {title} {\bibinfo {title} {The random walk's guide
  to anomalous diffusion: {A} fractional dynamics approach},\ }\href@noop {}
  {\bibfield  {journal} {\bibinfo  {journal} {Phys. Rep.}\ }\textbf {\bibinfo
  {volume} {339}},\ \bibinfo {pages} {1} (\bibinfo {year}
  {2000}{\natexlab{a}})}\BibitemShut {NoStop}%
\bibitem [{\citenamefont {Magdziarz}\ \emph {et~al.}(2008)\citenamefont
  {Magdziarz}, \citenamefont {Weron},\ and\ \citenamefont
  {Klafter}}]{MagdziarzWeronKlafter:2008}%
  \BibitemOpen
  \bibfield  {author} {\bibinfo {author} {\bibfnamefont {M.}~\bibnamefont
  {Magdziarz}}, \bibinfo {author} {\bibfnamefont {A.}~\bibnamefont {Weron}},\
  and\ \bibinfo {author} {\bibfnamefont {J.}~\bibnamefont {Klafter}},\
  }\bibfield  {title} {\bibinfo {title} {Equivalence of the fractional
  {F}okker-{P}lanck and subordinated {L}angevin equations: {T}he case of a
  time-dependent force},\ }\href@noop {} {\bibfield  {journal} {\bibinfo
  {journal} {Phys. Rev. Lett.}\ }\textbf {\bibinfo {volume} {101}},\ \bibinfo
  {pages} {210601} (\bibinfo {year} {2008})}\BibitemShut {NoStop}%
\bibitem [{\citenamefont {Eule}\ and\ \citenamefont
  {Friedrich}(2009)}]{EuleFriedrich:2009}%
  \BibitemOpen
  \bibfield  {author} {\bibinfo {author} {\bibfnamefont {S.}~\bibnamefont
  {Eule}}\ and\ \bibinfo {author} {\bibfnamefont {R.}~\bibnamefont
  {Friedrich}},\ }\bibfield  {title} {\bibinfo {title} {Subordinated {L}angevin
  equations for anomalous diffusion in external potentials-{B}iasing and
  decoupled external forces},\ }\href@noop {} {\bibfield  {journal} {\bibinfo
  {journal} {Europhys. Lett.}\ }\textbf {\bibinfo {volume} {86}},\ \bibinfo
  {pages} {30008} (\bibinfo {year} {2009})}\BibitemShut {NoStop}%
\bibitem [{\citenamefont {Cairoli}\ and\ \citenamefont
  {Baule}(2015)}]{CairoliBaule:2015}%
  \BibitemOpen
  \bibfield  {author} {\bibinfo {author} {\bibfnamefont {A.}~\bibnamefont
  {Cairoli}}\ and\ \bibinfo {author} {\bibfnamefont {A.}~\bibnamefont
  {Baule}},\ }\bibfield  {title} {\bibinfo {title} {Anomalous processes with
  general waiting times: {F}unctionals and multipoint structure},\ }\href@noop
  {} {\bibfield  {journal} {\bibinfo  {journal} {Phys. Rev. Lett.}\ }\textbf
  {\bibinfo {volume} {115}},\ \bibinfo {pages} {110601} (\bibinfo {year}
  {2015})}\BibitemShut {NoStop}%
\bibitem [{\citenamefont {Fedotov}\ and\ \citenamefont
  {Korabel}(2015)}]{FedotovKorabel:2015}%
  \BibitemOpen
  \bibfield  {author} {\bibinfo {author} {\bibfnamefont {S.}~\bibnamefont
  {Fedotov}}\ and\ \bibinfo {author} {\bibfnamefont {N.}~\bibnamefont
  {Korabel}},\ }\bibfield  {title} {\bibinfo {title} {Subdiffusion in an
  external potential: {A}nomalous effects hiding behind normal behavior},\
  }\href@noop {} {\bibfield  {journal} {\bibinfo  {journal} {Phys. Rev. E}\
  }\textbf {\bibinfo {volume} {91}},\ \bibinfo {pages} {042112} (\bibinfo
  {year} {2015})}\BibitemShut {NoStop}%
\bibitem [{\citenamefont {Kampen}(1992)}]{VanKampen:1992}%
  \BibitemOpen
  \bibfield  {author} {\bibinfo {author} {\bibfnamefont {N.~G.~V.}\
  \bibnamefont {Kampen}},\ }\href@noop {} {\emph {\bibinfo {title} {{Stochastic
  Processes in Physics and Chemistry}}}}\ (\bibinfo  {publisher}
  {North-Holland},\ \bibinfo {address} {Amsterdam},\ \bibinfo {year}
  {1992})\BibitemShut {NoStop}%
\bibitem [{\citenamefont {Coffey}\ \emph {et~al.}(2004)\citenamefont {Coffey},
  \citenamefont {Kalmykov},\ and\ \citenamefont
  {Waldron}}]{CoffeyKalmykovWaldron:2004}%
  \BibitemOpen
  \bibfield  {author} {\bibinfo {author} {\bibfnamefont {W.~T.}\ \bibnamefont
  {Coffey}}, \bibinfo {author} {\bibfnamefont {Y.~P.}\ \bibnamefont
  {Kalmykov}},\ and\ \bibinfo {author} {\bibfnamefont {J.~T.}\ \bibnamefont
  {Waldron}},\ }\href@noop {} {\emph {\bibinfo {title} {{The {L}angevin
  Equation}}}}\ (\bibinfo  {publisher} {World Scientific},\ \bibinfo {address}
  {Singapore},\ \bibinfo {year} {2004})\BibitemShut {NoStop}%
\bibitem [{\citenamefont {Wang}\ \emph {et~al.}(2009)\citenamefont {Wang},
  \citenamefont {Anthony}, \citenamefont {Bae},\ and\ \citenamefont
  {Granick}}]{WangAnthonyBaeGranick:2009}%
  \BibitemOpen
  \bibfield  {author} {\bibinfo {author} {\bibfnamefont {B.}~\bibnamefont
  {Wang}}, \bibinfo {author} {\bibfnamefont {S.~M.}\ \bibnamefont {Anthony}},
  \bibinfo {author} {\bibfnamefont {S.~C.}\ \bibnamefont {Bae}},\ and\ \bibinfo
  {author} {\bibfnamefont {S.}~\bibnamefont {Granick}},\ }\bibfield  {title}
  {\bibinfo {title} {Anomalous yet {B}rownian},\ }\href@noop {} {\bibfield
  {journal} {\bibinfo  {journal} {Proc. Natl. Acad. Sci. U.S.A.}\ }\textbf
  {\bibinfo {volume} {106}},\ \bibinfo {pages} {15160} (\bibinfo {year}
  {2009})}\BibitemShut {NoStop}%
\bibitem [{\citenamefont {Toyota}\ \emph {et~al.}(2011)\citenamefont {Toyota},
  \citenamefont {Head}, \citenamefont {Schmidt},\ and\ \citenamefont
  {Mizuno}}]{ToyotaHeadSchmidtMizuno:2011}%
  \BibitemOpen
  \bibfield  {author} {\bibinfo {author} {\bibfnamefont {T.}~\bibnamefont
  {Toyota}}, \bibinfo {author} {\bibfnamefont {D.~A.}\ \bibnamefont {Head}},
  \bibinfo {author} {\bibfnamefont {C.~F.}\ \bibnamefont {Schmidt}},\ and\
  \bibinfo {author} {\bibfnamefont {D.}~\bibnamefont {Mizuno}},\ }\bibfield
  {title} {\bibinfo {title} {Non-{G}aussian athermal fluctuations in active
  gels},\ }\href@noop {} {\bibfield  {journal} {\bibinfo  {journal} {Soft
  Matter}\ }\textbf {\bibinfo {volume} {7}},\ \bibinfo {pages} {3234} (\bibinfo
  {year} {2011})}\BibitemShut {NoStop}%
\bibitem [{\citenamefont {{Soares e Silva}}\ \emph {et~al.}(2014)\citenamefont
  {{Soares e Silva}}, \citenamefont {Stuhrmann}, \citenamefont {Betz},\ and\
  \citenamefont {Koenderink}}]{SilvaStuhrmannBetzKoenderink:2014}%
  \BibitemOpen
  \bibfield  {author} {\bibinfo {author} {\bibfnamefont {M.}~\bibnamefont
  {{Soares e Silva}}}, \bibinfo {author} {\bibfnamefont {B.}~\bibnamefont
  {Stuhrmann}}, \bibinfo {author} {\bibfnamefont {T.}~\bibnamefont {Betz}},\
  and\ \bibinfo {author} {\bibfnamefont {G.~H.}\ \bibnamefont {Koenderink}},\
  }\bibfield  {title} {\bibinfo {title} {Time-resolved microrheology of
  actively remodeling actomyos in networks},\ }\href@noop {} {\bibfield
  {journal} {\bibinfo  {journal} {New J. Phys.}\ }\textbf {\bibinfo {volume}
  {16}},\ \bibinfo {pages} {075010} (\bibinfo {year} {2014})}\BibitemShut
  {NoStop}%
\bibitem [{\citenamefont {Bhattacharya}\ \emph {et~al.}(2013)\citenamefont
  {Bhattacharya}, \citenamefont {Sharma}, \citenamefont {Saurabh},
  \citenamefont {De}, \citenamefont {Sain}, \citenamefont {Nandi},\ and\
  \citenamefont {Chowdhury}}]{Bhattacharya-etal:2013}%
  \BibitemOpen
  \bibfield  {author} {\bibinfo {author} {\bibfnamefont {S.}~\bibnamefont
  {Bhattacharya}}, \bibinfo {author} {\bibfnamefont {D.~K.}\ \bibnamefont
  {Sharma}}, \bibinfo {author} {\bibfnamefont {S.}~\bibnamefont {Saurabh}},
  \bibinfo {author} {\bibfnamefont {S.}~\bibnamefont {De}}, \bibinfo {author}
  {\bibfnamefont {A.}~\bibnamefont {Sain}}, \bibinfo {author} {\bibfnamefont
  {A.}~\bibnamefont {Nandi}},\ and\ \bibinfo {author} {\bibfnamefont
  {A.}~\bibnamefont {Chowdhury}},\ }\bibfield  {title} {\bibinfo {title}
  {{Plasticization of Poly(vinylpyrrolidone) thin films under ambient humidity:
  Insight from single-molecule tracer diffusion dynamics}},\ }\href@noop {}
  {\bibfield  {journal} {\bibinfo  {journal} {J. Phys. Chem. B}\ }\textbf
  {\bibinfo {volume} {117}},\ \bibinfo {pages} {7771} (\bibinfo {year}
  {2013})}\BibitemShut {NoStop}%
\bibitem [{\citenamefont {Kim}\ \emph {et~al.}(2013)\citenamefont {Kim},
  \citenamefont {Kim},\ and\ \citenamefont {Sung}}]{KimKimSung:2013}%
  \BibitemOpen
  \bibfield  {author} {\bibinfo {author} {\bibfnamefont {J.}~\bibnamefont
  {Kim}}, \bibinfo {author} {\bibfnamefont {C.}~\bibnamefont {Kim}},\ and\
  \bibinfo {author} {\bibfnamefont {B.~J.}\ \bibnamefont {Sung}},\ }\bibfield
  {title} {\bibinfo {title} {Simulation study of seemingly {F}ickian but
  heterogeneous dynamics of two dimensional colloids},\ }\href@noop {}
  {\bibfield  {journal} {\bibinfo  {journal} {Phys. Rev. Lett.}\ }\textbf
  {\bibinfo {volume} {110}},\ \bibinfo {pages} {047801} (\bibinfo {year}
  {2013})}\BibitemShut {NoStop}%
\bibitem [{\citenamefont {Beck}(2001)}]{Beck:2001}%
  \BibitemOpen
  \bibfield  {author} {\bibinfo {author} {\bibfnamefont {C.}~\bibnamefont
  {Beck}},\ }\bibfield  {title} {\bibinfo {title} {Dynamical foundations of
  nonextensive statistical mechanics},\ }\href@noop {} {\bibfield  {journal}
  {\bibinfo  {journal} {Phys. Rev. Lett.}\ }\textbf {\bibinfo {volume} {87}},\
  \bibinfo {pages} {180601} (\bibinfo {year} {2001})}\BibitemShut {NoStop}%
\bibitem [{\citenamefont {Beck}\ and\ \citenamefont
  {Cohen}(2003)}]{BeckCohen:2003}%
  \BibitemOpen
  \bibfield  {author} {\bibinfo {author} {\bibfnamefont {C.}~\bibnamefont
  {Beck}}\ and\ \bibinfo {author} {\bibfnamefont {E.~G.~D.}\ \bibnamefont
  {Cohen}},\ }\bibfield  {title} {\bibinfo {title} {Superstatistics},\
  }\href@noop {} {\bibfield  {journal} {\bibinfo  {journal} {Physica A}\
  }\textbf {\bibinfo {volume} {322}},\ \bibinfo {pages} {267} (\bibinfo {year}
  {2003})}\BibitemShut {NoStop}%
\bibitem [{\citenamefont {Beck}(2006)}]{Beck:2006}%
  \BibitemOpen
  \bibfield  {author} {\bibinfo {author} {\bibfnamefont {C.}~\bibnamefont
  {Beck}},\ }\bibfield  {title} {\bibinfo {title} {Superstatistical {B}rownian
  motion},\ }\href@noop {} {\bibfield  {journal} {\bibinfo  {journal} {Prog.
  Theor. Phys. Suppl.}\ }\textbf {\bibinfo {volume} {162}},\ \bibinfo {pages}
  {29} (\bibinfo {year} {2006})}\BibitemShut {NoStop}%
\bibitem [{\citenamefont {Wang}\ \emph {et~al.}(2012)\citenamefont {Wang},
  \citenamefont {Kuo}, \citenamefont {Bae},\ and\ \citenamefont
  {Granick}}]{WangKuoBaeGranick:2012}%
  \BibitemOpen
  \bibfield  {author} {\bibinfo {author} {\bibfnamefont {B.}~\bibnamefont
  {Wang}}, \bibinfo {author} {\bibfnamefont {J.}~\bibnamefont {Kuo}}, \bibinfo
  {author} {\bibfnamefont {S.~C.}\ \bibnamefont {Bae}},\ and\ \bibinfo {author}
  {\bibfnamefont {S.}~\bibnamefont {Granick}},\ }\bibfield  {title} {\bibinfo
  {title} {When {B}rownian diffusion is not {G}aussian},\ }\href@noop {}
  {\bibfield  {journal} {\bibinfo  {journal} {Nat. Mater.}\ }\textbf {\bibinfo
  {volume} {11}},\ \bibinfo {pages} {481} (\bibinfo {year} {2012})}\BibitemShut
  {NoStop}%
\bibitem [{\citenamefont {Hapca}\ \emph {et~al.}(2009)\citenamefont {Hapca},
  \citenamefont {Crawford},\ and\ \citenamefont
  {Young}}]{HapcaCrawfordYoung:2009}%
  \BibitemOpen
  \bibfield  {author} {\bibinfo {author} {\bibfnamefont {S.}~\bibnamefont
  {Hapca}}, \bibinfo {author} {\bibfnamefont {J.~W.}\ \bibnamefont
  {Crawford}},\ and\ \bibinfo {author} {\bibfnamefont {I.~M.}\ \bibnamefont
  {Young}},\ }\bibfield  {title} {\bibinfo {title} {Anomalous diffusion of
  heterogeneous populations characterized by normal diffusion at the individual
  level},\ }\href@noop {} {\bibfield  {journal} {\bibinfo  {journal} {J. R.
  Soc. Interface}\ }\textbf {\bibinfo {volume} {6}},\ \bibinfo {pages} {111}
  (\bibinfo {year} {2009})}\BibitemShut {NoStop}%
\bibitem [{\citenamefont {Barkai}\ and\ \citenamefont
  {Burov}(2020)}]{BarkaiBurov:2020}%
  \BibitemOpen
  \bibfield  {author} {\bibinfo {author} {\bibfnamefont {E.}~\bibnamefont
  {Barkai}}\ and\ \bibinfo {author} {\bibfnamefont {S.}~\bibnamefont {Burov}},\
  }\bibfield  {title} {\bibinfo {title} {Packets of diffusing particles exhibit
  universal exponential tails},\ }\href@noop {} {\bibfield  {journal} {\bibinfo
   {journal} {Phys. Rev. Lett.}\ }\textbf {\bibinfo {volume} {124}},\ \bibinfo
  {pages} {060603} (\bibinfo {year} {2020})}\BibitemShut {NoStop}%
\bibitem [{\citenamefont {Wang}\ \emph
  {et~al.}(2020{\natexlab{a}})\citenamefont {Wang}, \citenamefont {Barkai},\
  and\ \citenamefont {Burov}}]{WangBarkaiBurov:2020}%
  \BibitemOpen
  \bibfield  {author} {\bibinfo {author} {\bibfnamefont {W.~L.}\ \bibnamefont
  {Wang}}, \bibinfo {author} {\bibfnamefont {E.}~\bibnamefont {Barkai}},\ and\
  \bibinfo {author} {\bibfnamefont {S.}~\bibnamefont {Burov}},\ }\bibfield
  {title} {\bibinfo {title} {Large deviations for continuous time random
  walks},\ }\href@noop {} {\bibfield  {journal} {\bibinfo  {journal} {Entropy}\
  }\textbf {\bibinfo {volume} {22}},\ \bibinfo {pages} {697} (\bibinfo {year}
  {2020}{\natexlab{a}})}\BibitemShut {NoStop}%
\bibitem [{\citenamefont {Chubynsky}\ and\ \citenamefont
  {Slater}(2014)}]{ChubynskySlater:2014}%
  \BibitemOpen
  \bibfield  {author} {\bibinfo {author} {\bibfnamefont {M.~V.}\ \bibnamefont
  {Chubynsky}}\ and\ \bibinfo {author} {\bibfnamefont {G.~W.}\ \bibnamefont
  {Slater}},\ }\bibfield  {title} {\bibinfo {title} {Diffusing diffusivity: {A}
  model for anomalous, yet {B}rownian, diffusion},\ }\href@noop {} {\bibfield
  {journal} {\bibinfo  {journal} {Phys. Rev. Lett.}\ }\textbf {\bibinfo
  {volume} {113}},\ \bibinfo {pages} {098302} (\bibinfo {year}
  {2014})}\BibitemShut {NoStop}%
\bibitem [{\citenamefont {Chechkin}\ \emph {et~al.}(2017)\citenamefont
  {Chechkin}, \citenamefont {Seno}, \citenamefont {Metzler},\ and\
  \citenamefont {Sokolov}}]{ChechkinSenoMetzlerSokolov:2017}%
  \BibitemOpen
  \bibfield  {author} {\bibinfo {author} {\bibfnamefont {A.~V.}\ \bibnamefont
  {Chechkin}}, \bibinfo {author} {\bibfnamefont {F.}~\bibnamefont {Seno}},
  \bibinfo {author} {\bibfnamefont {R.}~\bibnamefont {Metzler}},\ and\ \bibinfo
  {author} {\bibfnamefont {I.~M.}\ \bibnamefont {Sokolov}},\ }\bibfield
  {title} {\bibinfo {title} {Brownian yet non-{G}aussian diffusion: {F}rom
  superstatistics to subordination of diffusing diffusivities},\ }\href@noop {}
  {\bibfield  {journal} {\bibinfo  {journal} {Phys. Rev. X}\ }\textbf {\bibinfo
  {volume} {7}},\ \bibinfo {pages} {021002} (\bibinfo {year}
  {2017})}\BibitemShut {NoStop}%
\bibitem [{\citenamefont {{\'{S}}l{\c{e}}zak}\ \emph
  {et~al.}(2018)\citenamefont {{\'{S}}l{\c{e}}zak}, \citenamefont {Metzler},\
  and\ \citenamefont {Magdziarz}}]{SlezakMetzlerMagdziarz:2018}%
  \BibitemOpen
  \bibfield  {author} {\bibinfo {author} {\bibfnamefont {J.}~\bibnamefont
  {{\'{S}}l{\c{e}}zak}}, \bibinfo {author} {\bibfnamefont {R.}~\bibnamefont
  {Metzler}},\ and\ \bibinfo {author} {\bibfnamefont {M.}~\bibnamefont
  {Magdziarz}},\ }\bibfield  {title} {\bibinfo {title} {Superstatistical
  generalised {L}angevin equation: {N}on-{G}aussian viscoelastic anomalous
  diffusion},\ }\href@noop {} {\bibfield  {journal} {\bibinfo  {journal} {New
  J. Phys.}\ }\textbf {\bibinfo {volume} {20}},\ \bibinfo {pages} {023026}
  (\bibinfo {year} {2018})}\BibitemShut {NoStop}%
\bibitem [{\citenamefont {Vitali}\ \emph {et~al.}(2018)\citenamefont {Vitali},
  \citenamefont {Sposini}, \citenamefont {Sliusarenko}, \citenamefont
  {Paradisi}, \citenamefont {Castellani},\ and\ \citenamefont
  {Pagnini}}]{Vitali.etal:2018}%
  \BibitemOpen
  \bibfield  {author} {\bibinfo {author} {\bibfnamefont {S.}~\bibnamefont
  {Vitali}}, \bibinfo {author} {\bibfnamefont {V.}~\bibnamefont {Sposini}},
  \bibinfo {author} {\bibfnamefont {O.}~\bibnamefont {Sliusarenko}}, \bibinfo
  {author} {\bibfnamefont {P.}~\bibnamefont {Paradisi}}, \bibinfo {author}
  {\bibfnamefont {G.}~\bibnamefont {Castellani}},\ and\ \bibinfo {author}
  {\bibfnamefont {G.}~\bibnamefont {Pagnini}},\ }\bibfield  {title} {\bibinfo
  {title} {Langevin equation in complex media and anomalous diffusion},\
  }\href@noop {} {\bibfield  {journal} {\bibinfo  {journal} {J. R. Soc.
  Interface}\ }\textbf {\bibinfo {volume} {15}},\ \bibinfo {pages} {20180282}
  (\bibinfo {year} {2018})}\BibitemShut {NoStop}%
\bibitem [{\citenamefont {Chen}\ and\ \citenamefont
  {Wang}(2021)}]{ChenWang:2021}%
  \BibitemOpen
  \bibfield  {author} {\bibinfo {author} {\bibfnamefont {Y.}~\bibnamefont
  {Chen}}\ and\ \bibinfo {author} {\bibfnamefont {X.~D.}\ \bibnamefont
  {Wang}},\ }\bibfield  {title} {\bibinfo {title} {Novel anomalous diffusion
  phenomena of underdamped langevin equation with random parameters},\
  }\href@noop {} {\bibfield  {journal} {\bibinfo  {journal} {New J. Phys.}\
  }\textbf {\bibinfo {volume} {23}},\ \bibinfo {pages} {123024} (\bibinfo
  {year} {2021})}\BibitemShut {NoStop}%
\bibitem [{\citenamefont {Sposini}\ \emph {et~al.}(2018)\citenamefont
  {Sposini}, \citenamefont {Chechkin}, \citenamefont {Seno}, \citenamefont
  {Pagnini},\ and\ \citenamefont
  {Metzler}}]{SposiniChechkinSenoPagniniMetzler:2018}%
  \BibitemOpen
  \bibfield  {author} {\bibinfo {author} {\bibfnamefont {V.}~\bibnamefont
  {Sposini}}, \bibinfo {author} {\bibfnamefont {A.~V.}\ \bibnamefont
  {Chechkin}}, \bibinfo {author} {\bibfnamefont {F.}~\bibnamefont {Seno}},
  \bibinfo {author} {\bibfnamefont {G.}~\bibnamefont {Pagnini}},\ and\ \bibinfo
  {author} {\bibfnamefont {R.}~\bibnamefont {Metzler}},\ }\bibfield  {title}
  {\bibinfo {title} {Random diffusivity from stochastic equations: {C}omparison
  of two models for {B}rownian yet non-{G}aussian diffusion},\ }\href@noop {}
  {\bibfield  {journal} {\bibinfo  {journal} {New J. Phys.}\ }\textbf {\bibinfo
  {volume} {20}},\ \bibinfo {pages} {043044} (\bibinfo {year}
  {2018})}\BibitemShut {NoStop}%
\bibitem [{\citenamefont {Jain}\ and\ \citenamefont
  {Sebastian}(2018)}]{JainSebastian:2018}%
  \BibitemOpen
  \bibfield  {author} {\bibinfo {author} {\bibfnamefont {R.}~\bibnamefont
  {Jain}}\ and\ \bibinfo {author} {\bibfnamefont {K.~L.}\ \bibnamefont
  {Sebastian}},\ }\bibfield  {title} {\bibinfo {title} {Diffusing diffusivity:
  {F}ractional {B}rownian oscillator model for subdiffusion and its solution},\
  }\href@noop {} {\bibfield  {journal} {\bibinfo  {journal} {Phys. Rev. E}\
  }\textbf {\bibinfo {volume} {98}},\ \bibinfo {pages} {052138} (\bibinfo
  {year} {2018})}\BibitemShut {NoStop}%
\bibitem [{\citenamefont {Ma\'{c}ka{\l}a}\ and\ \citenamefont
  {Magdziarz}(2019)}]{MackalaMagdziarz:2019}%
  \BibitemOpen
  \bibfield  {author} {\bibinfo {author} {\bibfnamefont {A.}~\bibnamefont
  {Ma\'{c}ka{\l}a}}\ and\ \bibinfo {author} {\bibfnamefont {M.}~\bibnamefont
  {Magdziarz}},\ }\bibfield  {title} {\bibinfo {title} {Statistical analysis of
  superstatistical fractional {B}rownian motion and applications},\ }\href@noop
  {} {\bibfield  {journal} {\bibinfo  {journal} {Phys. Rev. E}\ }\textbf
  {\bibinfo {volume} {99}},\ \bibinfo {pages} {012143} (\bibinfo {year}
  {2019})}\BibitemShut {NoStop}%
\bibitem [{\citenamefont {Wang}\ \emph
  {et~al.}(2020{\natexlab{b}})\citenamefont {Wang}, \citenamefont {Cherstvy},
  \citenamefont {Chechkin}, \citenamefont {Thapa}, \citenamefont {Seno},
  \citenamefont {Liu},\ and\ \citenamefont {Metzler}}]{Wang-etal:2020}%
  \BibitemOpen
  \bibfield  {author} {\bibinfo {author} {\bibfnamefont {W.}~\bibnamefont
  {Wang}}, \bibinfo {author} {\bibfnamefont {A.~G.}\ \bibnamefont {Cherstvy}},
  \bibinfo {author} {\bibfnamefont {A.~V.}\ \bibnamefont {Chechkin}}, \bibinfo
  {author} {\bibfnamefont {S.}~\bibnamefont {Thapa}}, \bibinfo {author}
  {\bibfnamefont {F.}~\bibnamefont {Seno}}, \bibinfo {author} {\bibfnamefont
  {X.}~\bibnamefont {Liu}},\ and\ \bibinfo {author} {\bibfnamefont
  {R.}~\bibnamefont {Metzler}},\ }\bibfield  {title} {\bibinfo {title}
  {Fractional {B}rownian motion with random diffusivity: {E}merging residual
  nonergodicity below the correlation time},\ }\href@noop {} {\bibfield
  {journal} {\bibinfo  {journal} {J. Phys. A}\ }\textbf {\bibinfo {volume}
  {53}},\ \bibinfo {pages} {474001} (\bibinfo {year}
  {2020}{\natexlab{b}})}\BibitemShut {NoStop}%
\bibitem [{\citenamefont {Wang}\ \emph
  {et~al.}(2020{\natexlab{c}})\citenamefont {Wang}, \citenamefont {Cherstvy},
  \citenamefont {Liu},\ and\ \citenamefont {Metzler}}]{Wang-etal:2020-2}%
  \BibitemOpen
  \bibfield  {author} {\bibinfo {author} {\bibfnamefont {W.}~\bibnamefont
  {Wang}}, \bibinfo {author} {\bibfnamefont {A.~G.}\ \bibnamefont {Cherstvy}},
  \bibinfo {author} {\bibfnamefont {X.}~\bibnamefont {Liu}},\ and\ \bibinfo
  {author} {\bibfnamefont {R.}~\bibnamefont {Metzler}},\ }\bibfield  {title}
  {\bibinfo {title} {Anomalous diffusion and nonergodicity for heterogeneous
  diffusion processes with fractional {G}aussian noise},\ }\href@noop {}
  {\bibfield  {journal} {\bibinfo  {journal} {Phys. Rev. E}\ }\textbf {\bibinfo
  {volume} {102}},\ \bibinfo {pages} {474001} (\bibinfo {year}
  {2020}{\natexlab{c}})}\BibitemShut {NoStop}%
\bibitem [{\citenamefont {Cherstvy}\ and\ \citenamefont
  {Metzler}(2016)}]{CherstvyMetzler:2016}%
  \BibitemOpen
  \bibfield  {author} {\bibinfo {author} {\bibfnamefont {A.~G.}\ \bibnamefont
  {Cherstvy}}\ and\ \bibinfo {author} {\bibfnamefont {R.}~\bibnamefont
  {Metzler}},\ }\bibfield  {title} {\bibinfo {title} {Anomalous diffusion in
  time-fluctuating non-stationary diffusivity landscapes},\ }\href@noop {}
  {\bibfield  {journal} {\bibinfo  {journal} {Phys. Chem. Chem. Phys.}\
  }\textbf {\bibinfo {volume} {18}},\ \bibinfo {pages} {23840} (\bibinfo {year}
  {2016})}\BibitemShut {NoStop}%
\bibitem [{\citenamefont {Wang}\ and\ \citenamefont
  {Chen}(2021)}]{WangChen:2021}%
  \BibitemOpen
  \bibfield  {author} {\bibinfo {author} {\bibfnamefont {X.~D.}\ \bibnamefont
  {Wang}}\ and\ \bibinfo {author} {\bibfnamefont {Y.}~\bibnamefont {Chen}},\
  }\bibfield  {title} {\bibinfo {title} {Ergodic property of {L}angevin systems
  with superstatistical, uncorrelated or correlated diffusivity},\ }\href@noop
  {} {\bibfield  {journal} {\bibinfo  {journal} {Physica A}\ }\textbf {\bibinfo
  {volume} {577}},\ \bibinfo {pages} {126090} (\bibinfo {year}
  {2021})}\BibitemShut {NoStop}%
\bibitem [{\citenamefont {Wang}\ and\ \citenamefont
  {Chen}(2022)}]{WangChen:2022}%
  \BibitemOpen
  \bibfield  {author} {\bibinfo {author} {\bibfnamefont {X.~D.}\ \bibnamefont
  {Wang}}\ and\ \bibinfo {author} {\bibfnamefont {Y.}~\bibnamefont {Chen}},\
  }\bibfield  {title} {\bibinfo {title} {Ergodic property of random diffusivity
  system with trapping events},\ }\href@noop {} {\bibfield  {journal} {\bibinfo
   {journal} {Phys. Rev. E}\ }\textbf {\bibinfo {volume} {105}},\ \bibinfo
  {pages} {014106} (\bibinfo {year} {2022})}\BibitemShut {NoStop}%
\bibitem [{\citenamefont {Kubo}(1966)}]{Kubo:1966}%
  \BibitemOpen
  \bibfield  {author} {\bibinfo {author} {\bibfnamefont {R.}~\bibnamefont
  {Kubo}},\ }\bibfield  {title} {\bibinfo {title} {The fluctuation-dissipation
  theorem},\ }\href@noop {} {\bibfield  {journal} {\bibinfo  {journal} {Rep.
  Prog. Phys.}\ }\textbf {\bibinfo {volume} {29}},\ \bibinfo {pages} {255}
  (\bibinfo {year} {1966})}\BibitemShut {NoStop}%
\bibitem [{\citenamefont {Marconi}\ \emph {et~al.}(2008)\citenamefont
  {Marconi}, \citenamefont {Puglisi}, \citenamefont {Rondoni},\ and\
  \citenamefont {Vulpiani}}]{MarconiPuglisiRondoniVulpiani:2008}%
  \BibitemOpen
  \bibfield  {author} {\bibinfo {author} {\bibfnamefont {U.~M.~B.}\
  \bibnamefont {Marconi}}, \bibinfo {author} {\bibfnamefont {A.}~\bibnamefont
  {Puglisi}}, \bibinfo {author} {\bibfnamefont {L.}~\bibnamefont {Rondoni}},\
  and\ \bibinfo {author} {\bibfnamefont {A.}~\bibnamefont {Vulpiani}},\
  }\bibfield  {title} {\bibinfo {title} {Fluctuation-dissipation: {R}esponse
  theory in statistical physics},\ }\href@noop {} {\bibfield  {journal}
  {\bibinfo  {journal} {Phys. Rep.}\ }\textbf {\bibinfo {volume} {461}},\
  \bibinfo {pages} {111} (\bibinfo {year} {2008})}\BibitemShut {NoStop}%
\bibitem [{\citenamefont {Barkai}\ and\ \citenamefont
  {Fleurov}(1998)}]{BarkaiFleurov:1998}%
  \BibitemOpen
  \bibfield  {author} {\bibinfo {author} {\bibfnamefont {E.}~\bibnamefont
  {Barkai}}\ and\ \bibinfo {author} {\bibfnamefont {V.~N.}\ \bibnamefont
  {Fleurov}},\ }\bibfield  {title} {\bibinfo {title} {Generalized {E}instein
  relation: {A} stochastic modeling approach},\ }\href@noop {} {\bibfield
  {journal} {\bibinfo  {journal} {Phys. Rev. E}\ }\textbf {\bibinfo {volume}
  {58}},\ \bibinfo {pages} {1296} (\bibinfo {year} {1998})}\BibitemShut
  {NoStop}%
\bibitem [{\citenamefont {B\'{e}nichou}\ and\ \citenamefont
  {Oshanin}(2002)}]{BenichouOshanin:2002}%
  \BibitemOpen
  \bibfield  {author} {\bibinfo {author} {\bibfnamefont {O.}~\bibnamefont
  {B\'{e}nichou}}\ and\ \bibinfo {author} {\bibfnamefont {G.}~\bibnamefont
  {Oshanin}},\ }\bibfield  {title} {\bibinfo {title} {Ultraslow
  vacancy-mediated tracer diffusion in two dimensions: {T}he {E}instein
  relation verified},\ }\href@noop {} {\bibfield  {journal} {\bibinfo
  {journal} {Phys. Rev. E}\ }\textbf {\bibinfo {volume} {66}},\ \bibinfo
  {pages} {031101} (\bibinfo {year} {2002})}\BibitemShut {NoStop}%
\bibitem [{\citenamefont {Shemer}\ and\ \citenamefont
  {Barkai}(2009)}]{ShemerBarkai:2009}%
  \BibitemOpen
  \bibfield  {author} {\bibinfo {author} {\bibfnamefont {Z.}~\bibnamefont
  {Shemer}}\ and\ \bibinfo {author} {\bibfnamefont {E.}~\bibnamefont
  {Barkai}},\ }\bibfield  {title} {\bibinfo {title} {Einstein relation and
  effective temperature for systems with quenched disorder},\ }\href@noop {}
  {\bibfield  {journal} {\bibinfo  {journal} {Phys. Rev. E}\ }\textbf {\bibinfo
  {volume} {80}},\ \bibinfo {pages} {031108} (\bibinfo {year}
  {2009})}\BibitemShut {NoStop}%
\bibitem [{\citenamefont {Froemberg}\ and\ \citenamefont
  {Barkai}(2013)}]{FroembergBarkai:2013-3}%
  \BibitemOpen
  \bibfield  {author} {\bibinfo {author} {\bibfnamefont {D.}~\bibnamefont
  {Froemberg}}\ and\ \bibinfo {author} {\bibfnamefont {E.}~\bibnamefont
  {Barkai}},\ }\bibfield  {title} {\bibinfo {title} {No-go theorem for
  ergodicity and an {E}instein relation},\ }\href@noop {} {\bibfield  {journal}
  {\bibinfo  {journal} {Phys. Rev. E}\ }\textbf {\bibinfo {volume} {88}},\
  \bibinfo {pages} {024101} (\bibinfo {year} {2013})}\BibitemShut {NoStop}%
\bibitem [{\citenamefont {Kubo}\ \emph {et~al.}(1985)\citenamefont {Kubo},
  \citenamefont {Toda},\ and\ \citenamefont
  {Hashitsume}}]{KuboTodaHashitsume:1985}%
  \BibitemOpen
  \bibfield  {author} {\bibinfo {author} {\bibfnamefont {R.}~\bibnamefont
  {Kubo}}, \bibinfo {author} {\bibfnamefont {M.}~\bibnamefont {Toda}},\ and\
  \bibinfo {author} {\bibfnamefont {N.}~\bibnamefont {Hashitsume}},\
  }\href@noop {} {\emph {\bibinfo {title} {{Statistical Physics II,
  Nonequilibrium Statistical Mechanics}}}}\ (\bibinfo  {publisher}
  {Springer-Verlag},\ \bibinfo {address} {Berlin},\ \bibinfo {year}
  {1985})\BibitemShut {NoStop}%
\bibitem [{\citenamefont {Zwanzig}(2001)}]{Zwanzig:2001}%
  \BibitemOpen
  \bibfield  {author} {\bibinfo {author} {\bibfnamefont {R.}~\bibnamefont
  {Zwanzig}},\ }\href@noop {} {\emph {\bibinfo {title} {{Non-Equilibrium
  Statistical Mechanics}}}}\ (\bibinfo  {publisher} {Oxford University Press},\
  \bibinfo {address} {New York},\ \bibinfo {year} {2001})\BibitemShut {NoStop}%
\bibitem [{\citenamefont {Wang}\ \emph
  {et~al.}(2019{\natexlab{a}})\citenamefont {Wang}, \citenamefont {Chen},\ and\
  \citenamefont {Deng}}]{WangChenDeng:2019}%
  \BibitemOpen
  \bibfield  {author} {\bibinfo {author} {\bibfnamefont {X.~D.}\ \bibnamefont
  {Wang}}, \bibinfo {author} {\bibfnamefont {Y.}~\bibnamefont {Chen}},\ and\
  \bibinfo {author} {\bibfnamefont {W.~H.}\ \bibnamefont {Deng}},\ }\bibfield
  {title} {\bibinfo {title} {L{\'{e}}vy-walk-like {L}angevin dynamics},\
  }\href@noop {} {\bibfield  {journal} {\bibinfo  {journal} {New J. Phys.}\
  }\textbf {\bibinfo {volume} {21}},\ \bibinfo {pages} {013024} (\bibinfo
  {year} {2019}{\natexlab{a}})}\BibitemShut {NoStop}%
\bibitem [{\citenamefont {Wang}\ \emph
  {et~al.}(2019{\natexlab{b}})\citenamefont {Wang}, \citenamefont {Deng},\ and\
  \citenamefont {Chen}}]{WangDengChen:2019}%
  \BibitemOpen
  \bibfield  {author} {\bibinfo {author} {\bibfnamefont {X.~D.}\ \bibnamefont
  {Wang}}, \bibinfo {author} {\bibfnamefont {W.~H.}\ \bibnamefont {Deng}},\
  and\ \bibinfo {author} {\bibfnamefont {Y.}~\bibnamefont {Chen}},\ }\bibfield
  {title} {\bibinfo {title} {Ergodic properties of heterogeneous diffusion
  processes in a potential well},\ }\href@noop {} {\bibfield  {journal}
  {\bibinfo  {journal} {J. Chem. Phys.}\ }\textbf {\bibinfo {volume} {150}},\
  \bibinfo {pages} {164121} (\bibinfo {year} {2019}{\natexlab{b}})}\BibitemShut
  {NoStop}%
\bibitem [{\citenamefont {Wang}\ \emph {et~al.}(2018)\citenamefont {Wang},
  \citenamefont {Chen},\ and\ \citenamefont {Deng}}]{WangChenDeng:2018}%
  \BibitemOpen
  \bibfield  {author} {\bibinfo {author} {\bibfnamefont {X.~D.}\ \bibnamefont
  {Wang}}, \bibinfo {author} {\bibfnamefont {Y.}~\bibnamefont {Chen}},\ and\
  \bibinfo {author} {\bibfnamefont {W.~H.}\ \bibnamefont {Deng}},\ }\bibfield
  {title} {\bibinfo {title} {Feynman-{K}ac equation revisited},\ }\href@noop {}
  {\bibfield  {journal} {\bibinfo  {journal} {Phys. Rev. E}\ }\textbf {\bibinfo
  {volume} {98}},\ \bibinfo {pages} {052114} (\bibinfo {year}
  {2018})}\BibitemShut {NoStop}%
\bibitem [{\citenamefont {Cairoli}\ and\ \citenamefont
  {Baule}(2017)}]{CairoliBaule:2017}%
  \BibitemOpen
  \bibfield  {author} {\bibinfo {author} {\bibfnamefont {A.}~\bibnamefont
  {Cairoli}}\ and\ \bibinfo {author} {\bibfnamefont {A.}~\bibnamefont
  {Baule}},\ }\bibfield  {title} {\bibinfo {title} {Feynman-{K}ac equation for
  anomalous processes with space- and time-dependent forces},\ }\href@noop {}
  {\bibfield  {journal} {\bibinfo  {journal} {J. Phys. A}\ }\textbf {\bibinfo
  {volume} {50}},\ \bibinfo {pages} {164002} (\bibinfo {year}
  {2017})}\BibitemShut {NoStop}%
\bibitem [{\citenamefont {Metzler}\ and\ \citenamefont
  {Klafter}(2000{\natexlab{b}})}]{MetzlerKlafter:00}%
  \BibitemOpen
  \bibfield  {author} {\bibinfo {author} {\bibfnamefont {R.}~\bibnamefont
  {Metzler}}\ and\ \bibinfo {author} {\bibfnamefont {J.}~\bibnamefont
  {Klafter}},\ }\bibfield  {title} {\bibinfo {title} {The random walk's guide
  to anomalous diffusion: a fractional dynamics approach},\ }\href@noop {}
  {\bibfield  {journal} {\bibinfo  {journal} {Phys. Rep.}\ }\textbf {\bibinfo
  {volume} {339}},\ \bibinfo {pages} {1} (\bibinfo {year}
  {2000}{\natexlab{b}})}\BibitemShut {NoStop}%
\bibitem [{\citenamefont {Friedrich}\ \emph {et~al.}(2006)\citenamefont
  {Friedrich}, \citenamefont {Jenko}, \citenamefont {Baule},\ and\
  \citenamefont {Eule}}]{FriedrichJenkoBauleEule:2006}%
  \BibitemOpen
  \bibfield  {author} {\bibinfo {author} {\bibfnamefont {R.}~\bibnamefont
  {Friedrich}}, \bibinfo {author} {\bibfnamefont {F.}~\bibnamefont {Jenko}},
  \bibinfo {author} {\bibfnamefont {A.}~\bibnamefont {Baule}},\ and\ \bibinfo
  {author} {\bibfnamefont {S.}~\bibnamefont {Eule}},\ }\bibfield  {title}
  {\bibinfo {title} {Anomalous diffusion of inertial, weakly damped
  particles},\ }\href@noop {} {\bibfield  {journal} {\bibinfo  {journal} {Phys.
  Rev. Lett.}\ }\textbf {\bibinfo {volume} {96}},\ \bibinfo {pages} {230601}
  (\bibinfo {year} {2006})}\BibitemShut {NoStop}%
\bibitem [{\citenamefont {Turgeman}\ \emph {et~al.}(2009)\citenamefont
  {Turgeman}, \citenamefont {Carmi},\ and\ \citenamefont
  {Barkai}}]{TurgemanCarmiBarkai:2009}%
  \BibitemOpen
  \bibfield  {author} {\bibinfo {author} {\bibfnamefont {L.}~\bibnamefont
  {Turgeman}}, \bibinfo {author} {\bibfnamefont {S.}~\bibnamefont {Carmi}},\
  and\ \bibinfo {author} {\bibfnamefont {E.}~\bibnamefont {Barkai}},\
  }\bibfield  {title} {\bibinfo {title} {Fractional {F}eynman-{K}ac equation
  for non-{B}rownian functionals},\ }\href@noop {} {\bibfield  {journal}
  {\bibinfo  {journal} {Phys. Rev. Lett.}\ }\textbf {\bibinfo {volume} {103}},\
  \bibinfo {pages} {190201} (\bibinfo {year} {2009})}\BibitemShut {NoStop}%
\bibitem [{\citenamefont {Koszto{\l}owicz}\ and\ \citenamefont
  {Dutkiewicz}(2021)}]{KosztolowiczDutkiewicz:2021}%
  \BibitemOpen
  \bibfield  {author} {\bibinfo {author} {\bibfnamefont {T.}~\bibnamefont
  {Koszto{\l}owicz}}\ and\ \bibinfo {author} {\bibfnamefont {A.}~\bibnamefont
  {Dutkiewicz}},\ }\bibfield  {title} {\bibinfo {title} {Subdiffusion equation
  with caputo fractional derivative with respect to another function},\
  }\href@noop {} {\bibfield  {journal} {\bibinfo  {journal} {Phys. Rev. E}\
  }\textbf {\bibinfo {volume} {104}},\ \bibinfo {pages} {014118} (\bibinfo
  {year} {2021})}\BibitemShut {NoStop}%
\bibitem [{\citenamefont {Fogedby}(1994)}]{Fogedby:1994}%
  \BibitemOpen
  \bibfield  {author} {\bibinfo {author} {\bibfnamefont {H.~C.}\ \bibnamefont
  {Fogedby}},\ }\bibfield  {title} {\bibinfo {title} {Langevin equations for
  continuous time {L}\'{e}vy flights},\ }\href@noop {} {\bibfield  {journal}
  {\bibinfo  {journal} {Phys. Rev. E}\ }\textbf {\bibinfo {volume} {50}},\
  \bibinfo {pages} {1657} (\bibinfo {year} {1994})}\BibitemShut {NoStop}%
\bibitem [{\citenamefont {Metzler}\ and\ \citenamefont
  {Klafter}(2000{\natexlab{c}})}]{MetzlerKlafter:2000-2}%
  \BibitemOpen
  \bibfield  {author} {\bibinfo {author} {\bibfnamefont {R.}~\bibnamefont
  {Metzler}}\ and\ \bibinfo {author} {\bibfnamefont {J.}~\bibnamefont
  {Klafter}},\ }\bibfield  {title} {\bibinfo {title} {From a generalized
  {C}hapman-{K}olmogorov equation to the fractional {K}lein-{K}ramers
  equation},\ }\href@noop {} {\bibfield  {journal} {\bibinfo  {journal} {J.
  Phys. Chem. B}\ }\textbf {\bibinfo {volume} {104}},\ \bibinfo {pages} {3851}
  (\bibinfo {year} {2000}{\natexlab{c}})}\BibitemShut {NoStop}%
\bibitem [{\citenamefont {Eule}\ \emph {et~al.}(2012)\citenamefont {Eule},
  \citenamefont {Zaburdaev}, \citenamefont {Friedrich},\ and\ \citenamefont
  {Geisel}}]{EuleZaburdaevFriedrichGeisel:2012}%
  \BibitemOpen
  \bibfield  {author} {\bibinfo {author} {\bibfnamefont {S.}~\bibnamefont
  {Eule}}, \bibinfo {author} {\bibfnamefont {V.}~\bibnamefont {Zaburdaev}},
  \bibinfo {author} {\bibfnamefont {R.}~\bibnamefont {Friedrich}},\ and\
  \bibinfo {author} {\bibfnamefont {T.}~\bibnamefont {Geisel}},\ }\bibfield
  {title} {\bibinfo {title} {Langevin description of superdiffusive {L}\'{e}vy
  processes},\ }\href@noop {} {\bibfield  {journal} {\bibinfo  {journal} {Phys.
  Rev. E}\ }\textbf {\bibinfo {volume} {86}},\ \bibinfo {pages} {041134}
  (\bibinfo {year} {2012})}\BibitemShut {NoStop}%
\bibitem [{\citenamefont {Risken}(1989)}]{Risken:1989}%
  \BibitemOpen
  \bibfield  {author} {\bibinfo {author} {\bibfnamefont {H.}~\bibnamefont
  {Risken}},\ }\href@noop {} {\emph {\bibinfo {title} {The {F}okker-{P}lanck
  {E}quation}}}\ (\bibinfo  {publisher} {Springer-Verlag},\ \bibinfo {address}
  {Berlin},\ \bibinfo {year} {1989})\BibitemShut {NoStop}%
\bibitem [{\citenamefont {Denisov}\ \emph {et~al.}(2009)\citenamefont
  {Denisov}, \citenamefont {Horsthemke},\ and\ \citenamefont
  {H{\"{a}}nggi}}]{DenisovHorsthemkeHanggi:2009}%
  \BibitemOpen
  \bibfield  {author} {\bibinfo {author} {\bibfnamefont {S.~I.}\ \bibnamefont
  {Denisov}}, \bibinfo {author} {\bibfnamefont {W.}~\bibnamefont
  {Horsthemke}},\ and\ \bibinfo {author} {\bibfnamefont {P.}~\bibnamefont
  {H{\"{a}}nggi}},\ }\bibfield  {title} {\bibinfo {title} {Generalized
  {F}okker-{P}lanck equation: {D}erivation and exact solutions},\ }\href@noop
  {} {\bibfield  {journal} {\bibinfo  {journal} {Eur. Phys. J. B}\ }\textbf
  {\bibinfo {volume} {68}},\ \bibinfo {pages} {567} (\bibinfo {year}
  {2009})}\BibitemShut {NoStop}%
\bibitem [{\citenamefont {Cairoli}\ \emph {et~al.}(2018)\citenamefont
  {Cairoli}, \citenamefont {Klages},\ and\ \citenamefont
  {Baule}}]{CairoliKlagesBaule:2018}%
  \BibitemOpen
  \bibfield  {author} {\bibinfo {author} {\bibfnamefont {A.}~\bibnamefont
  {Cairoli}}, \bibinfo {author} {\bibfnamefont {R.}~\bibnamefont {Klages}},\
  and\ \bibinfo {author} {\bibfnamefont {A.}~\bibnamefont {Baule}},\ }\bibfield
   {title} {\bibinfo {title} {Weak {G}alilean invariance as a selection
  principle for coarse-grained diffusive models},\ }\href@noop {} {\bibfield
  {journal} {\bibinfo  {journal} {Proc. Natl. Acad. Sci. USA}\ }\textbf
  {\bibinfo {volume} {115}},\ \bibinfo {pages} {5714} (\bibinfo {year}
  {2018})}\BibitemShut {NoStop}%
\bibitem [{\citenamefont {Chen}\ \emph
  {et~al.}(2019{\natexlab{a}})\citenamefont {Chen}, \citenamefont {Wang},\ and\
  \citenamefont {Deng}}]{ChenWangDeng:2019-2}%
  \BibitemOpen
  \bibfield  {author} {\bibinfo {author} {\bibfnamefont {Y.}~\bibnamefont
  {Chen}}, \bibinfo {author} {\bibfnamefont {X.~D.}\ \bibnamefont {Wang}},\
  and\ \bibinfo {author} {\bibfnamefont {W.~H.}\ \bibnamefont {Deng}},\
  }\bibfield  {title} {\bibinfo {title} {Subdiffusion in an external force
  field},\ }\href@noop {} {\bibfield  {journal} {\bibinfo  {journal} {Phys.
  Rev. E}\ }\textbf {\bibinfo {volume} {99}},\ \bibinfo {pages} {042125}
  (\bibinfo {year} {2019}{\natexlab{a}})}\BibitemShut {NoStop}%
\bibitem [{\citenamefont {Gradshteyn}\ \emph {et~al.}(1980)\citenamefont
  {Gradshteyn}, \citenamefont {Ryzhik}, \citenamefont {Geraniums},\ and\
  \citenamefont {Tseytlin}}]{GradshteynRyzhikGeraniumsTseytlin:1980}%
  \BibitemOpen
  \bibfield  {author} {\bibinfo {author} {\bibfnamefont {I.~S.}\ \bibnamefont
  {Gradshteyn}}, \bibinfo {author} {\bibfnamefont {I.~M.}\ \bibnamefont
  {Ryzhik}}, \bibinfo {author} {\bibfnamefont {Y.~V.}\ \bibnamefont
  {Geraniums}},\ and\ \bibinfo {author} {\bibfnamefont {M.~Y.}\ \bibnamefont
  {Tseytlin}},\ }\href@noop {} {\emph {\bibinfo {title} {{Table of Integrals,
  Series, and Products}}}}\ (\bibinfo  {publisher} {Academic Press},\ \bibinfo
  {address} {USA},\ \bibinfo {year} {1980})\BibitemShut {NoStop}%
\bibitem [{\citenamefont {Dankel}(1991)}]{Dankel:1991}%
  \BibitemOpen
  \bibfield  {author} {\bibinfo {author} {\bibfnamefont {T.}~\bibnamefont
  {Dankel}},\ }\bibfield  {title} {\bibinfo {title} {On the distribution of the
  integrated square of the {O}rnstein-{U}hlenbeck process},\ }\href@noop {}
  {\bibfield  {journal} {\bibinfo  {journal} {SIAM J. Appl. Math.}\ }\textbf
  {\bibinfo {volume} {51}},\ \bibinfo {pages} {568} (\bibinfo {year}
  {1991})}\BibitemShut {NoStop}%
\bibitem [{\citenamefont {Chen}\ \emph
  {et~al.}(2019{\natexlab{b}})\citenamefont {Chen}, \citenamefont {Wang},\ and\
  \citenamefont {Deng}}]{ChenWangDeng:2019-3}%
  \BibitemOpen
  \bibfield  {author} {\bibinfo {author} {\bibfnamefont {Y.}~\bibnamefont
  {Chen}}, \bibinfo {author} {\bibfnamefont {X.~D.}\ \bibnamefont {Wang}},\
  and\ \bibinfo {author} {\bibfnamefont {W.~H.}\ \bibnamefont {Deng}},\
  }\bibfield  {title} {\bibinfo {title} {Langevin picture of {L}\'{e}vy walk in
  a constant force field},\ }\href@noop {} {\bibfield  {journal} {\bibinfo
  {journal} {Phys. Rev. E}\ }\textbf {\bibinfo {volume} {100}},\ \bibinfo
  {pages} {062141} (\bibinfo {year} {2019}{\natexlab{b}})}\BibitemShut
  {NoStop}%
\end{thebibliography}%

\end{document}